\pdfoutput=1
\documentclass[acmsmall]{acmart}
\usepackage[T1]{fontenc}
\usepackage{enumitem}
\usepackage{amsmath,amsfonts}
\usepackage{algorithmic}
\usepackage{array}
\usepackage[caption=false,font=normalsize,labelfont=sf,textfont=sf]{subfig}
\usepackage{textcomp}
\usepackage{stfloats}
\usepackage{url}
\usepackage{verbatim}
\usepackage{graphicx}
\usepackage{balance}
\usepackage{adjustbox}
\usepackage{booktabs}
\usepackage{multirow}
\usepackage{threeparttable}
\usepackage{arydshln}
\usepackage{tcolorbox}
\usepackage{color}
\usepackage{soul}
\usepackage{fontawesome}
\usepackage{makecell}
\usepackage{natbib}
\usepackage{colortbl}
\usepackage{siunitx}
\usepackage{microtype}
\definecolor{mygray}{gray}{.9}
\usepackage{tikz}
\usepackage[T1]{fontenc}
\usepackage{aecompl}

\usepackage{hyperref}
\usepackage{graphicx}
\usepackage{subfig}
\usepackage{subcaption}

\NewDocumentCommand{\framecolorbox}{oommm}
 {
  \IfValueTF{#1}
   {\IfValueTF{#2}
    {\fcolorbox{#3}{#4}{\makebox[#1][#2]{#5}}}
    {\fcolorbox{#3}{#4}{\makebox[#1]{#5}}}%
   }
   {\fcolorbox{#3}{#4}{#5}}%
 }

\AtBeginDocument{
  \providecommand\BibTeX{{
    \normalfont B\kern-0.5em{\scshape i\kern-0.25em b}\kern-0.8em\TeX}}}

\AtBeginDocument{
  \providecommand\BibTeX{{
    \normalfont B\kern-0.5em{\scshape i\kern-0.25em b}\kern-0.8em\TeX}}}

\begin{document}

\title{Large Language Models for Multi-Lingual Equivalent Mutant Detection: An Extended Empirical Study}

\author{Honglin Shu}
\authornote{These authors contributed equally to this work.}
\affiliation{%
  \institution{Tianjin University}
  \city{Tianjin}
  \country{China}
  ~and
  \institution{Kyushu University}
  \city{Fukuoka}
  \country{Japan}
  }
\email{shu.honglin.167@s.kyushu-u.ac.jp}

\author{Zhao Tian}
\authornotemark[1]
\affiliation{
  \institution{College of Intelligence and 
  Computing, Tianjin University}
  \city{Tianjin}
  \country{China}
}
\email{tianzhao@tju.edu.cn}

\author{Dong Wang}
\authornote{Corresponding Author}
\affiliation{%
  \institution{College of Intelligence and 
  Computing, Tianjin University}
  \city{Tianjin}
  \country{China}
}
\email{dong_w@tju.edu.cn}

\author{Junji Yu}
\affiliation{%
  \institution{College of Intelligence and 
  Computing, Tianjin University}
  \city{Tianjin}
  \country{China}
}
\email{junjiyu@tju.edu.cn}

\author{Jiazhe Zhang}
\affiliation{%
  \institution{College of Intelligence and 
  Computing, Tianjin University}
  \city{Tianjin}
  \country{China}
}
\email{2839197907z@gmail.com}

\author{Xuejie Cao}
\affiliation{%
  \institution{College of Intelligence and 
  Computing, Tianjin University}
  \city{Tianjin}
  \country{China}
}
\email{caoxuejie@tju.edu.cn}

\author{Junjie Chen}
\affiliation{
  \institution{College of Intelligence and 
  Computing, Tianjin University}
  \city{Tianjin}
  \country{China}
}
\email{junjiechen@tju.edu.cn}

\author{Yasutaka Kamei}
\affiliation{%
  \institution{Kyushu University}
  \city{Fukuoka}
  \country{Japan}
}
\email{kamei@ait.kyushu-u.ac.jp}

\renewcommand{\shortauthors}{Wang et al.}

\begin{abstract}
Mutation testing is a powerful technique for ensuring software quality. 
However, the presence of equivalent mutants introduces unnecessary costs and biases, limiting its practical effectiveness.
Although numerous equivalent mutant detection (EMD) methods have been proposed, they often face distinct challenges: pure-code analysis methods can be limited by their reliance on specific compiler infrastructures, while existing machine-learning approaches remain constrained by scarce training data and limited generalization to unseen mutants.
Large language models (LLMs) have recently demonstrated remarkable performance across diverse code-related tasks by better capturing program semantics. 
Yet their potential for EMD remains largely unexplored, particularly in the multi-lingual context. 
This paper presents the first comprehensive empirical study on LLMs for EMD, using 3,302 Java and 1,088 C mutant pairs to benchmark against state-of-the-art methods, explore strategy variations, assess efficiency, and evaluate cross-lingual generalization.
Experimental results show that LLM-based approaches achieve higher F1-scores than the evaluated traditional methods, with fine-tuned code embedding yielding the highest detection accuracy among the tested strategies.
Moreover, LLM-based approaches strike a practical balance between effectiveness and efficiency with inference times comparable to existing machine-learning models.
Importantly, fine-tuned LLMs demonstrate measurable generalization across programming languages.
These findings establish LLMs as a viable and efficient approach for tackling the longstanding challenge of equivalent mutant detection, offering new directions for advancing mutation testing in practice.
\end{abstract}

\ccsdesc[500]{Software and its engineering~Software testing and debugging}

\keywords{Equivalent Mutant Detection, Mutation Testing, Large Language Model, Empirical Study}

\maketitle

\section{Introduction}
\label{sec:introduction}
Mutation testing~\cite{andrews2005mutation, jia2010analysis} is a widely used technique for evaluating test suite effectiveness by introducing artificial faults into programs, known as \textit{mutants}, through small and targeted modifications to the original code.
Beyond measuring test effectiveness, mutation testing has been effectively applied to various software testing and debugging activities, including test case prioritization~\cite{lou2015mutation}, automated bug detection~\cite{pradel2018deepbugs}, and fault localization~\cite{papadakis2015metallaxis}, where it has consistently achieved state-of-the-art performance.

Despite its widespread adoption and significance, mutation testing faces a variety of practical hurdles during industrial deployment, including developer workflow integration and high computational costs. 
While recent industrial experience reports demonstrate that equivalent mutants are not the primary blocking factor for adoption in practice\mbox{~\cite{petrovic2018state, beller2021would, vercacmmen2023validation, van2024integrating}}, they nonetheless represent one of the critical issues in this domain\mbox{~\cite{papadakis2019mutation, titcheu2020selecting}}, a problem proven undecidable for over three decades.
Equivalent mutants are redundant variants that exhibit identical behavior to the original program across all possible test cases\mbox{~\cite{jia2010analysis, madeyski2013overcoming}}, with studies reporting their prevalence in real-world development between 4\% and 39\%\mbox{~\cite{madeyski2013overcoming}}. 
Their presence introduces significant bias into mutation-based analysis. The standard mutation score metric, calculated using only non-equivalent mutants, can never reach 100\% if these redundant variants are not properly filtered, leading developers to question the reliability of otherwise adequate test suites. 
While they are part of a broader set of industrial challenges, these cost and bias issues undermine the practical effectiveness of mutation testing, potentially delaying development and compromising software quality. 
As a result, effectively and efficiently detecting equivalent mutants remains an important research priority for improving the reliability of mutation analysis.

Over the years, researchers have proposed numerous equivalent mutant detection (EMD) techniques to mitigate the challenges posed by equivalent mutants~\cite{kintis2017detecting,gheyi2021identifying}.
Traditional methods primarily rely on pre-defined rules, such as constraint-based testing~\cite{offutt1997automatically,schuler2013covering,kushigian2019medusa,baer2020mutantdistiller} and compiler optimization~\cite{papadakis2015trivial,houshmand2017tce+,kintis2017detecting}.
While these pure-code analysis methods are foundational and highly beneficial because they do not require training data, their scope can sometimes be constrained by a reliance on specific compiler infrastructures or the need for manually crafted heuristics.
To complement these approaches, researchers have explored learning-based EMD techniques, including conventional machine learning classifiers such as KNN and SVM\mbox{~\cite{naeem2020machine,brito2020preliminary,chung2022augmenting}}, as well as tree-based neural network models\mbox{~\cite{peacock2021automatic}}.
While these approaches offer alternative ways to analyze code by leveraging extracted code features, they remain limited in capturing deep program semantics, particularly when distinguishing subtle syntactic variations between mutants. Moreover, their effectiveness is hindered by the scarcity of labeled equivalent mutant data and difficulties in generalizing to previously unseen mutant patterns.

Lately, large language models (LLMs) have exhibited remarkable capabilities in software engineering~\cite{schafer2023empirical,sallou2024breaking}, leveraging pre-trained corpora containing extensive code repositories (such as StarCoder~\cite{li2023starcoder} and Code Llama~\cite{touvron2023llama}) to acquire generalized knowledge that enhances various code-related tasks~\cite{yang2024empirical,liu2024your}.
Notably, LLMs have demonstrated significant potential across multiple software testing domains through fine-tuning and prompt engineering strategies, including test case generation~\cite{schafer2023empirical, yang2024enhancing}, program debugging~\cite{feng2024prompting, li2023nuances}, and program repair~\cite{huang2023empirical}.
The detection of equivalent mutants fundamentally requires deep semantic code understanding, with LLMs, having been pre-trained on diverse code repositories, possessing superior semantic comprehension capabilities compared to traditional learning-based EMD techniques that suffer from limited training data.
This advantage positions LLMs as promising solutions to the persistent data scarcity challenge in this domain.
Our first empirical study~\cite{tian2024large} confirms the superiority of LLMs for EMD, yet three critical limitations remain unaddressed:
(1) First, we exclusively evaluated performance on Java programs, without assessing other foundational languages such as C. 
However, LLMs may exhibit varying levels of sensitivity to high- and low-resource programming languages depending on their pre-training data~\cite{multilingualvul}, rendering conclusions drawn from a single language potentially non-generalizable across languages.
(2) Second, we examined only a limited set of LLMs. 
Given the rapid evolution of LLMs, the performance of more recent and advanced models, such as DeepSeek and Qwen, remains underexplored.
(3) Third, given the limited availability of EMD data in any single programming language, it is essential to explore whether LLMs trained on cross-language datasets can effectively generalize to specific target languages.

In this study, we extend the empirical analysis to 3,302 method-level Java mutant pairs and 1,088 C mutant pairs to comprehensively investigate how LLMs can be leveraged for equivalent mutant detection in the multi-lingual context.
We systematically evaluate 13 state-of-the-art LLMs on both Java and C environments.
To evaluate both the effectiveness and efficiency of LLM-based approaches and mitigate the three aforementioned limitations, we formulate the following five research questions:
\\
\noindent
    \textbf{- RQ1: What is the performance of state-of-the-art LLMs on equivalent mutant detection?}\\ 
    \mbox{\textit{\underline{Motivation:} }} To justify adopting LLMs for automated mutant filtering or developer warning systems, we must first determine if they fundamentally demonstrate higher effectiveness compared to conventional techniques in the critical metrics of precision, recall, and F1-score, ensuring they are viable for strict real-world deployment.\\
    \textit{\underline{Result:} } LLM-based techniques achieve higher F1-scores than all ten EMD baselines in equivalent mutant detection, achieving average F1-score improvements of 75.18\% and 557.79\% over Compiler-based techniques, 19.14\% and 7.21\% over ML-based techniques, and 12.75\% and 48.83\% over Tree-based NN techniques in Java and C environments, respectively.
    \\
    \textbf{- RQ2: What is the best strategy using LLMs for equivalent mutant detection?}\\
    \mbox{\textit{\underline{Motivation:} }} Because different engineering workflows prioritize either strict soundness to avoid false alarms or casting a wider net to minimize false negatives, identifying the most effective LLM architecture and utilization strategy is necessary to provide actionable configuration guidelines for test engineers. \\
    \textit{\underline{Result:} } Fine-tuned UniXCoder yields the highest F1-scores among all LLM combinations and strategies on Java environments, improving F1-scores by 1.16\%$\sim$84.10\%. Similarly, fine-tuned CodeT5+ demonstrates the highest measured effectiveness performance in C environments, with F1-score improvements of 0.08\%$\sim$471.96\% compared to alternative LLMs using different strategies. In contrast, LLMs relying exclusively on prompting strategies yield lower evaluation metrics.
    \\
    \textbf{- RQ3: What degree of orthogonality exists between our studied EMD techniques?}\\
    \mbox{\textit{\underline{Motivation:} }} In practical deployment, models may exhibit varying confidence across different perspectives such as mutation operators. Understanding this orthogonality allows test engineers to selectively apply automated removal only for high-confidence recommended operators, while defaulting to IDE developer warnings for more complex semantic mutations.\\
    \textit{\underline{Result:} } In both Java and C environments, the LLM-based techniques and fine-tuned code embedding strategy tend to achieve higher accuracy than alternative EMD categories and LLM strategies. 
    This capability is evident in their ability to make unique correct/incorrect detections and in their effectiveness across various mutation operators, which strongly reinforce our findings from RQ1 and RQ2.
    \\
    \textbf{- RQ4: How efficient are our studied EMD techniques?} \\
    \mbox{\textit{\underline{Motivation:} }} For an EMD technique to be viable as a real-time developer warning plugin in an IDE or as a fast gatekeeper in an automated testing pipeline, its inference time must be low enough to avoid unacceptable computational bottlenecks. \\
    \textit{\underline{Result:} } The inference time of the best-performing LLM-based technique (\SI{0.0431}{s} for Java and \SI{0.0215}{s} for C) is shorter than the best-performing Compiler-based technique (\SI{2.3537}{s} for Java and \SI{0.6613}{s} for C), while being slightly slower than the best-performing ML-based technique (\SI{0.0019}{s} for Java and \SI{0.0021}{s} for C) and Tree-based NN technique (\SI{0.0274}{s} for Java). These results demonstrate a practical trade-off between computational efficiency and effectiveness for LLMs.
    \\
    \textbf{- RQ5: How effective are fine-tuned LLMs at detecting cross-lingual equivalent mutants?} \\
    \mbox{\textit{\underline{Motivation:} }} Real-world industrial repositories frequently consist of polyglot codebases. Establishing cross-lingual effectiveness determines whether test engineers can deploy a single, unified EMD model across their entire pipeline, rather than bearing the heavy computational and maintenance costs overhead of training and maintaining separate, language-specific models. \\
    \textit{\underline{Result:} } Cross-lingual fine-tuning enhances LLMs' effectiveness in equivalent mutant detection across Java and C, with fine-tuning with instruction showing measurable benefits (3.55\%$\sim$3.70\% F1-score improvement), while fine-tuning code embedding performs worse under this approach.
    
\noindent
    \textbf{Contributions.} To sum up, the contributions of this study are:
\begin{itemize}[leftmargin=10pt]
    \item Our study presents the first comprehensive, large-scale empirical evaluation of LLMs for equivalent mutant detection in the multi-lingual context, by examining five key dimensions: effectiveness relative to existing EMD techniques, the impact of different LLM strategies, orthogonality between various EMD approaches, computational efficiency, and generalization of fine-tuned LLMs on cross-lingual EMD.

    \item The study demonstrates that LLM-based equivalent mutant detection techniques achieve higher F1-scores compared to traditional baselines, and illustrates their promising potential for cross-programming language environments. Furthermore, we introduce a qualitative analysis of unique incorrect detections to investigate the specific code structures where LLMs fail compared to conventional methods.
    
    \item Our research offers comprehensive insights into both the strengths and limitations of LLMs for equivalent mutant detection. These findings provide crucial guidance for researchers working to improve LLM-based approaches in mutation testing and broader software engineering applications. To promote reproducibility and further advancement in this field, we have made all data, code, and detailed analysis procedures publicly available~\cite{EMD2024}.
\end{itemize}

\smallskip
\noindent
\textbf{Paper Extension.} 
This paper extends our prior work~\cite{tian2024large}, published as a research paper at the 33rd ACM SIGSOFT International Symposium on Software Testing and Analysis (ISSTA) with the distinguished paper award.
The key differences can be summarized as follows:
(I) \textit{Extension of programming language.} We extend the programming language from Java to C for each research question to further explore the effectiveness of LLMs on EMD.
(II) \textit{Extension of studied LLMs.} To ensure experimental fairness and finding accuracy, we expanded the LLMs in our study. Specifically, we added three of the latest LLMs: DeepSeek-Coder, Llama 3, and Qwen2.5-Coder.
(III) \textit{Exploring the effectiveness of the studied LLM on cross-lingual scenarios.} To provide a more comprehensive perspective, we further explore the generalization ability of LLMs on EMD by training on a cross-language EMD dataset and inference on a single programming language.

\section{Background and Related Work}
\label{sec:background}

\subsection{Mutation Testing}
\label{subsec:mutation_testing}
Mutation testing is a sophisticated program analysis technique that systematically modifies source code to introduce potential defects~\cite{shi2019mitigating,moradi2023muakka}.
These modifications follow defined patterns called \textit{mutation operators}, which are developed based on syntactic rules derived from the target programming language's grammar~\cite{ojdanic2023mutation}.
For example, applying a relational operator replacement mutation operator might transform ``\texttt{if(x==y)}'' in the original program to ``\texttt{if(x!=y)}'', creating what's known as a mutant.
The fundamental premise of mutation testing is that these artificially introduced faults effectively simulate real-world defects~\cite{andrews2006using,gopinath2014mutations}.
The primary purpose of mutation testing is to assess test suite effectiveness~\cite{gopinath2018if,perretta2022use}. A mutant is considered ``killed'' when at least one test case detects it; otherwise, it remains ``live''.
The effectiveness of mutation testing is measured by the mutation score, the percentage of killed mutants. This metric is calculated by dividing the number of killed mutants (those causing test failures) by the total number of non-equivalent mutants.

\subsection{Equivalent Mutants}
\label{subsec:equiv_mut}
The equivalent mutant problem represents a critical challenge in mutation testing that researchers have investigated extensively for decades~\cite{kintis2017detecting,bartocci2023property}. 
A mutant is considered equivalent when it produces identical behavior to the original program across all possible test cases, despite having syntactic differences.
These semantically equivalent mutants cannot be killed by any test case, making them a major obstacle to the widespread adoption of mutation testing in industry. 
Their presence increases computational overhead and introduces substantial evaluation bias~\cite{yao2014study,papadakis2019mutation}.
Studies indicate that equivalent mutants constitute between 4\% and 39\% of all mutants in real-world software development~\cite{madeyski2013overcoming}. 
The proliferation of mutants not only increases computational demands and evaluation bias~\cite{holling2016nequivack}, but also necessitates considerable effort for equivalent mutant detection and classification.

Detecting equivalent mutants poses a significant challenge in practice due to the undecidable nature of program equivalence in code mutation~\cite{arcaini2017novel,kim2022predictive}. To address this issue, researchers have developed various EMD techniques.
Traditional EMD approaches include genetic algorithms~\cite{adamopoulos2004overcome}, constraint-based testing~\cite{offutt1997automatically,kushigian2019medusa,baer2020mutantdistiller}, coverage analysis~\cite{schuler2010covering,schuler2013covering,papadakis2013mutation,papadakis2014mitigating}, automata language equivalence~\cite{devroey2018model}, software behavior graphs~\cite{gong2022equivalent}, structural pattern matching~\cite{ems}, dynamic subsumption relations~\cite{guimaraes2020optimizing,gheyi2021identifying}, and compiler optimization~\cite{papadakis2015trivial,houshmand2017tce+,kintis2017detecting}.
More recent advancements have explored learning-based EMD techniques, incorporating conventional binary classifiers like KNN and SVM~\cite{naeem2020machine,brito2020preliminary,chung2022augmenting} as well as tree-based neural network (NN) models~\cite{peacock2021automatic}. Notably, initial evaluations of tree-based NN approaches showed promising results, achieving 90\% classification accuracy with 582 mutants generated from just two mutation operators~\cite{peacock2021automatic}.

Several existing EMD techniques extract features by executing mutant programs within test suites~\cite{baer2020mutantdistiller,naeem2020machine,chung2022augmenting,gong2022equivalent}. While test suite information can improve equivalent mutant detection accuracy, generating and executing numerous test cases requires substantial time and computational resources~\cite{jia2010analysis}. Therefore, this paper focuses exclusively on techniques that leverage the semantic learning of code, specifically examining how LLMs can be applied to equivalent mutant detection.

\subsection{Large Language Models}
\label{subsec:llm}
LLMs have emerged as a cornerstone of the NLP domain due to their remarkable performance capabilities, exemplified by models such as Llama 2~\cite{touvron2023llama} and PaLM 2~\cite{anil2023palm}.
Beyond general-purpose applications, specialized LLMs have been developed with training on code repositories to extend their text generation prowess to programming domains. Notable examples include StarCoder~\cite{li2023starcoder} and Code Llama~\cite{roziere2023code}, which demonstrate the versatility of these advanced AI systems.

Software testing with LLMs has experienced remarkable growth in recent years~\cite{Wang2023SoftwareTW}. These models have proven particularly effective for test case generation~\cite{schafer2023empirical,xia2023universal}, program debugging~\cite{feng2024prompting,li2023nuances}, and program repair~\cite{huang2023empirical}, leveraging approaches such as fine-tuning and prompt engineering.
In the domain of test case generation, \citet{schafer2023empirical} conducted a comprehensive empirical evaluation of LLMs for automatic unit test generation using various prompting strategies. Similarly, \citet{xia2023universal} introduced Fuzz4All, which employs LLMs as mutation engines to generate diverse and realistic inputs across multiple programming languages, surpassing traditional language-specific fuzzers in performance.
For program debugging, \citet{feng2024prompting} developed AdbGPT, a lightweight approach that reproduces bugs from bug reports through sophisticated prompt engineering. \citet{li2023nuances} created Differential Prompting, a technique that efficiently identifies failure-inducing test cases by leveraging compilable code synthesized from inferred intentions.
Regarding program repair, \citet{huang2023empirical} performed an extensive analysis of five popular LLMs' repair capabilities using fine-tuning techniques, demonstrating that LLM-based methods substantially outperform conventional automated program repair approaches.
Furthermore, LLMs have been applied to mutation testing~\cite{leam++,tip2025llmorpheus,harman2025mutation}. For instance, ~\citet{tip2025llmorpheus} used LLMs to replace the placeholders that are introduced at designated locations by the mutation testing tool.

The effectiveness of LLMs in equivalent mutant detection remains systematically explored despite previous efforts. 
While ~\citet{ma2023scope} conducted a preliminary study using ChatGPT on a limited dataset of 200 mutants and ~\citet{tian2024large} conducted an empirical study on a single programming language environment, our research addresses these limitations through a more comprehensive evaluation. We assess the performance of both typical and state-of-the-art LLMs for equivalent mutant detection, examining diverse aspects including detection strategies, orthogonality with existing techniques, computational efficiency, and cross-lingual generalization.
\section{Study Design}
\label{sec:evaluation_design}
\begin{figure}[t!]
    \centering
    \includegraphics[width=.8\linewidth]{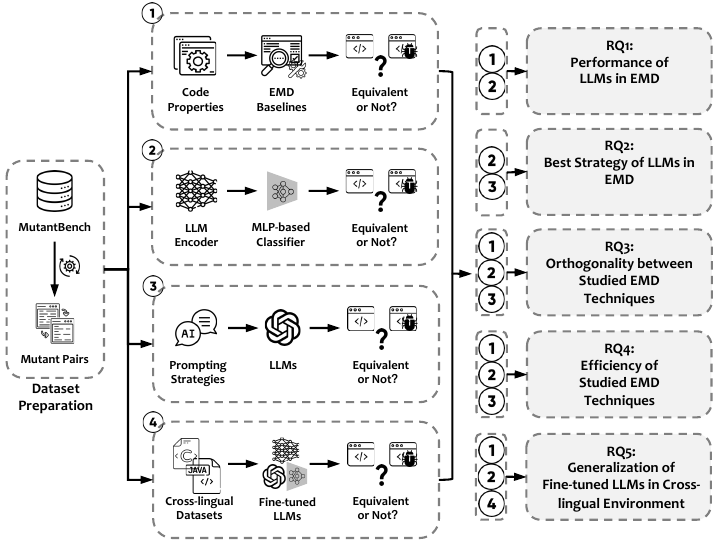}
    \caption{Overview of experimental design. \textcircled{1}/\textcircled{2}/\textcircled{3}/\textcircled{4} represent the workflow of EMD baselines/code embedding strategies/prompting strategies/cross\-lingual generalization, respectively.
    }
    \label{fig:overview}
\end{figure}

Figure~\ref{fig:overview} demonstrates the overview of our study design. 
Initially, we construct training and test datasets comprising code pairs of the original program and its mutant, based on the widely-used MutantBench~\cite{van2021mutantbench} dataset. 
Our research explores the effectiveness of existing EMD techniques and state-of-the-art LLMs on equivalent mutant detection (RQ1), and investigates how various strategies utilized by LLMs impact equivalent mutant detection performance (RQ2). 
We also analyze the orthogonality between different EMD techniques to understand their unique characteristics (RQ3). 
The study concludes with a quantitative measurement of time efficiency across all studied EMD techniques (RQ4). 
Finally, we explore the effectiveness of fine-tuned LLMs on cross-programming EMD environment (RQ5).

\subsection{Dataset Preparation}
\label{subsec:dataset}
\textbf{Studied dataset}. 
For performance evaluation of EMD techniques, we utilize MutantBench~\cite{van2021mutantbench}, the most widely-used benchmark containing 4,400 mutant pairs across C and Java programming languages.
MutantBench integrates multiple open-source datasets~\cite{yao2014study,kintis2016analysing, houshmand2017tce+, baer2020mutantdistiller}, providing enhanced diversity and covering a broader range of mutant types.
Our study concentrates on Java and C languages.
The dataset comprises 3,302 Java mutant pairs from 19 programs and 1,088 C mutant pairs from 18 programs, as shown in Table~\ref{tab:dataset}. 

\begin{table}[thp]
    \caption{Statistics of Java and C programs from MutantBench}
    \label{tab:dataset}
    \centering
    \tabcolsep=2.2mm
    \small
    \begin{adjustbox}{max width=1.0 \textwidth,center}
        \begin{tabular}{ clccc }
            \toprule
            \toprule
            \textbf{Language} & \textbf{Dataset} & \textbf{\#Programs} & \textbf{\#EQ} & \textbf{\#NEQ} \\ 
            \midrule
            \multirow{2}{*}{Java} & \textit{$MutantBench_{train-Java}$} & \multirow{2}{*}{19}  & 250 & 1,402 \\
            & \textit{$MutantBench_{test-Java}$}  &  &  249 & 1,401 \\
            \midrule
            \multirow{2}{*}{C} & \textit{$MutantBench_{train-C}$} & \multirow{2}{*}{18} & 454 & 90 \\
            & \textit{$MutantBench_{test-C}$}  &  &  453 & 91 \\
            \bottomrule
            \bottomrule
        \end{tabular}
    \end{adjustbox}
    \begin{tablenotes}
        \footnotesize
        \centering
        \item[*] \#Programs, \#Methods, \#EQ, and \#NEQ represent the number of programs, \\ 
        methods, equivalent mutants, and non-equivalent mutants, respectively.
    \end{tablenotes}
\end{table}

\textbf{Data pre-processing}. 
Our data pre-processing involved two key steps to accommodate LLM input requirements.
We narrowed our focus to method-level equivalence detection rather than program-level analysis. 
From a practical use-case perspective, this design choice can directly align with real-world developer workflows. 
First, mutation testing is primarily deployed as a unit-level testing activity, where test engineers design test suites to validate individual methods in isolation\mbox{~\cite{10.1145/2635868.2635929, jia2010analysis}}.
Because the individual method is the standard unit of verification in object-oriented unit testing, bounding the equivalence analysis to the method level directly mirrors these real-world testing workflows.
Second, many of the most widely used mutation testing frameworks (e.g., MuJava, Major, and PIT) generate and inject mutants specifically at the method-level. 
Because these method-level mutants are the standard artifacts produced and analyzed in practical environments, studying equivalence at this specific granularity is highly meaningful and directly applicable to existing industry standards.
Following established practices from previous research~\cite{tufano2019learning,tian2022learning}, we extracted the methods containing mutations from both original programs and their mutants.
This process yielded 3,302 method-level Java mutant pairs and 1,088 method-level C mutant pairs with all natural language comments removed.

\textbf{Construction of training and test datasets}.
Following established methodologies~\cite{cervellera2017distribution,lu2021variance}, we implemented stratified sampling to minimize bias and ensure representative training and test datasets.
We used Java as our example dataset, with C following an identical process.
Firstly, the 3,302 total mutant pairs were divided into two subgroups: $MutantBench_{eq}$, comprising solely equivalent mutant pairs, and $MutantBench_{neq}$, encompassing the rest of non-equivalent mutant pairs.
Subsequently, we randomly selected approximately ~50\% of the mutant pairs from both $MutantBench_{eq}$ and $MutantBench_{neq}$ to create a training dataset called $MutantBench_{train-Java}$. 
The remaining mutant pairs were combined to form our test dataset, $MutantBench_{test-Java}$.
Finally, we obtained 1,652 mutant pairs in $MutantBench_{train-Java}$, consisting of 250 equivalent mutant pairs and 1,402 non-equivalent mutant pairs. 
For $MutantBench_{test-Java}$, we ended up with 1,650 mutant pairs, comprising 249 equivalent mutant pairs and 1,401 non-equivalent mutant pairs.
For the C language, we obtained 545 mutant pairs in $MutantBench_{train-C}$, consisting of 454 equivalent and 90 non-equivalent mutant pairs. Similarly, we collected 545 mutant pairs for $MutantBench_{test-C}$, comprising 453 equivalent and 91 non-equivalent mutant pairs.
In particular, we confirmed that there is no data leakage between our training and test datasets through manual inspection.

\subsection{Experimental Large Language Models}
\label{subsec:exp_llm}
In our study, we investigated how 13 state-of-the-art LLMs perform in detecting equivalent mutants.
These models have gained widespread adoption in the literature~\cite{du2023pre, tian2023code, Wang2023SoftwareTW}, including:
\begin{itemize}[leftmargin=10pt]
    \item \textbf{CodeBERT}~\cite{feng2020codebert} is a prominent pre-trained model that leverages a multilayer Transformer architecture to process bimodal data encompassing both source code and natural languages.
    \item \textbf{GraphCodeBERT}~\cite{guo2020graphcodebert} is a transformer-based pre-trained model that incorporates semantic-level code information to obtain more effective code representations.

    \item \textbf{PLBART}~\cite{ahmad2021unified} leverages the BART architecture to create a bidirectional and autoregressive transformer model. It uses denoising pre-training objectives on a dataset that combines both source code and natural language.
    
    \item \textbf{CodeT5}~\cite{wang2021codet5} is an encoder-decoder model that understands code token types. Built on T5's architecture, it uses sequence-to-sequence pre-training with denoising objectives.

    \item \textbf{CodeT5+}~\cite{wang2023codet5+} extends CodeT5 with an architecture featuring a shallow encoder and deep decoder. Training occurs in phases, beginning with unimodal data before progressing to bimodal data.

    \item \textbf{UniXCoder}~\cite{guo2022unixcoder} leverages multi-modal information (including abstract syntax trees and code comments) to enhance code representation through a unified cross-modal pre-trained transformer-based model.
    
    \item \textbf{StarCoder}~\cite{li2023starcoder} builds upon the SantaCoder architecture and includes its dedicated encoder model named StarEncoder. It supports infilling capabilities and enables efficient large-batch inference through multi-query attention mechanisms.
    
    \item \textbf{Code Llama}~\cite{roziere2023code} stands as a leading LLM specifically designed for code generation and infilling, built upon the Llama 2 architecture. This decoder-only model underwent additional fine-tuning with 500B tokens drawn from a specialized code-focused dataset.

    \item \textbf{Llama 3}~\cite{grattafiori2024llama} is an open-source family of instruction-tuned language models, trained on an unprecedented around 15 trillion tokens and designed to excel at multilingual understanding, coding, reasoning, and long-context tasks (128K-token context window). 

    \item \textbf{Qwen2.5-Coder}~\cite{hui2024qwen2} is a specialized large language model for code tasks built on the Qwen2.5 architecture. It has undergone pre-training on over 5.5 trillion tokens extracted from public code repositories and web-crawled data containing code-related content. 

    \item \textbf{DeepSeek-Coder}~\cite{guo2024deepseek} is an open-source family of code-specific LLMs trained from scratch on 2 trillion tokens across 87 programming languages. These models utilize a fill-in-the-blank approach with a 16K context window to enhance both code generation and infilling capabilities.

    \item \textbf{Text-Embedding Models}~\cite{textemb2024} are advanced embedding models created by OpenAI~\cite{openai2024}. 
    These models generate high-quality text and code embeddings with superior representation capabilities. In our research, we utilized all three available versions: Text-Embedding-Ada-002, Text-Embedding-3-Small, and Text-Embedding-3-Large.
    
    \item \textbf{ChatGPT}~\cite{chatgpt2022} represents a significant advancement in LLMs, with potential to transform multiple domains including software engineering.
    The model is trained on extensive corpora of natural language and code, utilizing reinforcement learning from human feedback to enhance instruction following.
    In our study, we specifically examined two LLM versions: GPT-3.5-Turbo \cite{chatgpt2022} and GPT-4 \cite{openai2023gpt4}.
\end{itemize}
In summary, the studied LLMs can be categorized into two main architectural types: encoder-based LLMs and decoder-only LLMs.
Encoder-based LLMs encompass two subcategories: encoder-only models (i.e., CodeBERT, GraphCodeBERT, and Text-Embedding Models) and encoder-decoder models (i.e., PLBART, CodeT5, UniXCoder, CodeT5+, and StarCoder).
Decoder-only LLMs comprise models like Code Llama, Llama 3, Qwen2.5-Coder, DeepSeek-Coder, and ChatGPT.

\subsection{Pre-trained Language Models for Code Embedding}
\label{subsec:exp_plm}
Pre-trained encoder LLMs have demonstrated remarkable advancements in code classification tasks, including code clone detection~\cite{khajezade2024investigating}, functionality classification~\cite{zhang2019novel}, and vulnerability detection~\cite{ding2021towards}. 
These encoder LLMs acquire general-purpose code embedding knowledge through pre-training on vast repositories of code snippets. 
For deployment in various downstream tasks, researchers commonly employ a multilayer perceptron (MLP) classifier that leverages code embeddings generated by the pre-trained encoder LLMs to classify specific properties~\cite{guo2020graphcodebert,niu2023empirical,tian2023fly}. 
We designate this prevalent LLM methodology as the \textbf{pre-trained code embedding strategy}, which serves as a foundational LLM paradigm.

Therefore, we leveraged pre-trained code embeddings with our encoder LLMs to evaluate their effectiveness in equivalent mutant detection. 
Our approach uses an encoder-based classifier framework that combines an LLM encoder with an MLP-based classifier, as illustrated in Figure~\ref{fig:overview}.
The framework integrates pre-trained encoder LLMs to train domain-specific classifiers using our dataset. 
Following established practices~\cite{guo2020graphcodebert,niu2023empirical}, we maintained fixed parameters for the pre-trained LLM encoder while only updating the MLP-based classifier parameters during training. 
Through multiple training iterations, the classifier learns to detect equivalent mutants by utilizing the code embeddings produced by the pre-trained LLM encoder.

In particular, for encoder LLMs (i.e., CodeBERT, GraphCodeBERT, PLBART, CodeT5, UniXCoder, CodeT5+, StarCoder, Text-Embedding-Ada-002, Text-Embedding-3-Small, and Text-Embedding-3-Large), we used their encoder components to obtain embedding vectors.
However, exceptional decoder-only LLMs like Code Llama, Llama 3, Qwen2.5-Coder, Deepseek-Coder, and ChatGPT cannot use this pre-trained code embedding strategy.

\begin{figure}[]
    \centering
    \subfloat[\small A zero-shot prompting template for equivalent mutant detection]{
        \includegraphics[width=.7\linewidth]{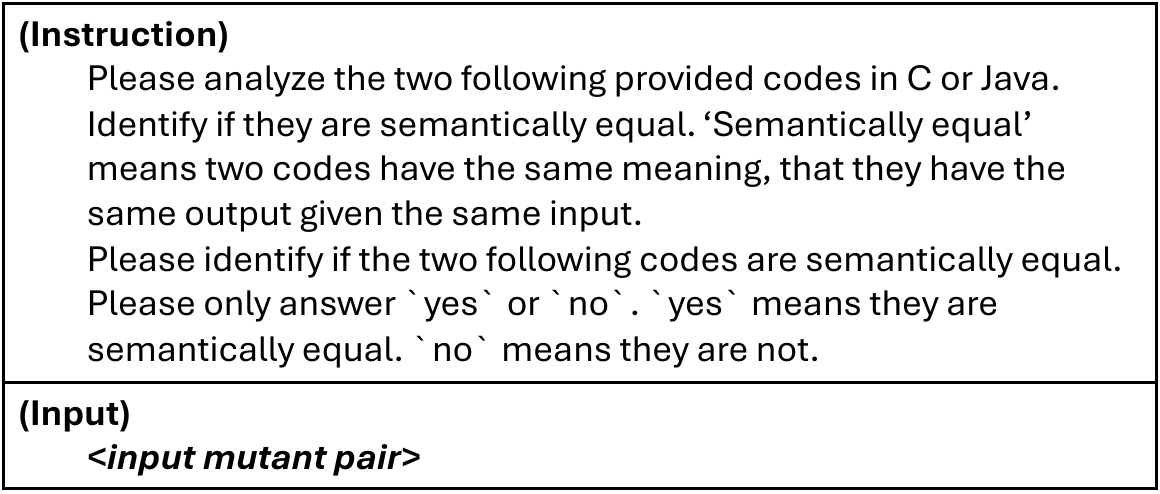}
        \label{fig:zsp_example}
    }\\
    \subfloat[\small A few-shot prompting template for equivalent mutant detection]{
        \includegraphics[width=.7\linewidth]{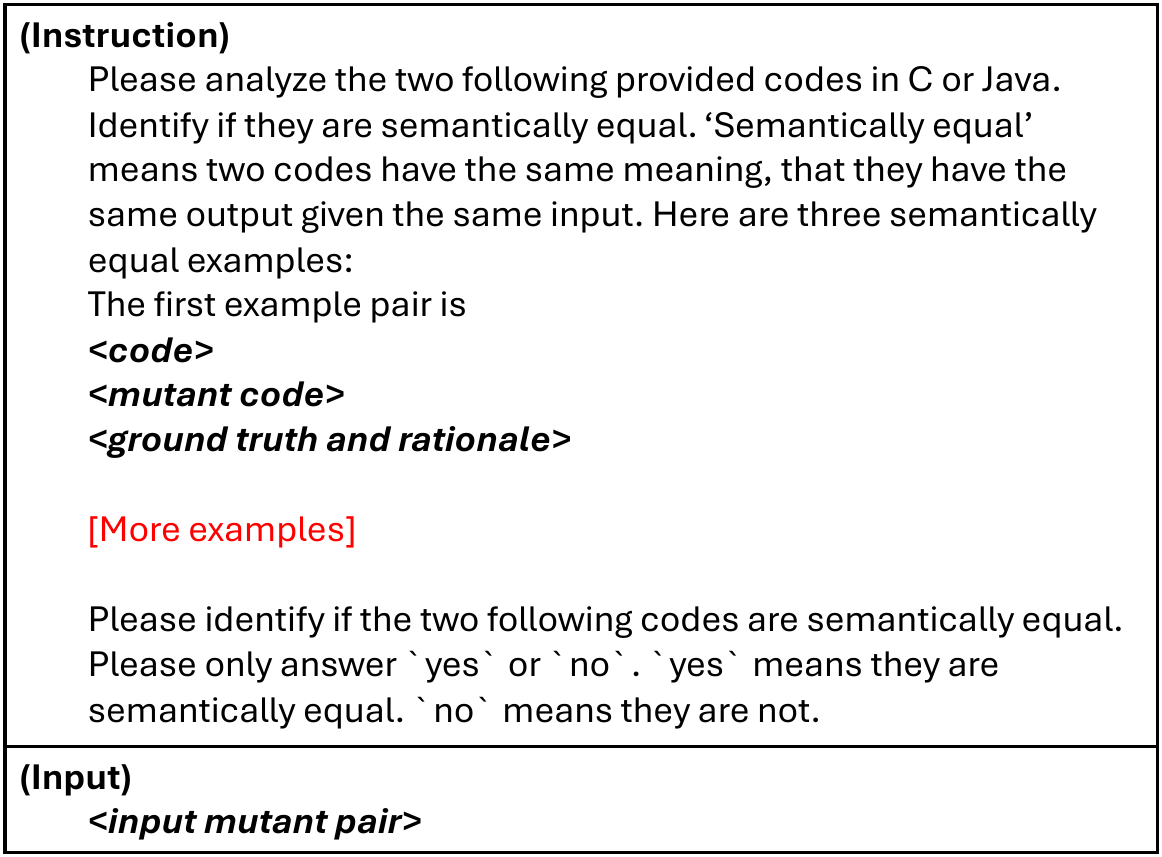}
        \label{fig:fsp_example}
    }
    \caption{Zero-shot prompt template and few-shot prompt template}
    \label{fig:prompt_example}
\end{figure}

\subsection{Strategies for Large Language Models}
\label{subsec:exp_stg}
This research explores how different LLM adaptation strategies affect equivalent mutant detection.
We selected four methods representing two complementary paradigms for LLM adaptation: training-time techniques that require parameter updates, and inference-time techniques that rely on prompt-based instructions.
We selected four methods representing two complementary paradigms for LLM adaptation: training-time techniques that require time-consuming parameter updates, and inference-time techniques that rely on prompt-based instructions without updating model parameters.
Within the training-time techniques, the \textbf{fine-tuned code embedding strategy} involves full parameter updating without natural language instructions, an approach proven to significantly improve task-specific performance\mbox{~\cite{liu2023pre, zhang2023revisiting}}.
The \textbf{fine-tuning with instruction strategy} bridges these paradigms by combining parameter updates with structured natural language alignment, which effectively aligns models with user intent and improves generalization to unseen tasks\mbox{~\cite{ouyang2022training, zhang2026instruction}}. 
On the inference-time side, \textbf{zero-shot} and \textbf{few-shot prompting strategies} utilize in-context learning; they require no weight updates, instead achieving adaptation through pre-trained knowledge and task-specific demonstrations\mbox{~\cite{kojima2022large}}. 
Together, these approaches robustly evaluate the trade-offs between parameter updates and prompt-based inference, leading to the development of our four specialized LLM strategies:

\begin{itemize}[leftmargin=10pt]
    \item \textbf{Fine-tuned code embedding strategy}: we maintained the same encoder-based classifier framework and hyper-parameter configurations used in the \textbf{pre-trained code embedding strategy} across all encoder LLMs. Unlike the previous approach, we did not freeze the encoder parameters but instead jointly optimized both the encoder and classifier parameters during the training process. 
    
    \item \textbf{Zero-shot prompting strategy}: we devised a prompt that doesn't use examples, instead directly employing a mutant pair with structured instructions (i.e., ``Please identify if the two above codes are semantically equal. Please only answer `yes' or `no'. `yes' means they are semantically equal. `no' means they are not.'') to evaluate LLMs for equivalent mutant detection.
    As shown in Figure\mbox{~\ref{fig:zsp_example}}, the zero-shot prompt template defines the instructions and input formats used to evaluate the LLMs.

    \item \textbf{Few-shot prompting strategy}: 
    it enables LLMs to understand the relationship between mutant pairs and semantic equivalence by learning from randomly selected examples in the format $<$\textit{mutant pair}, \textit{equivalence}$>$. 
    The method combines these demonstration examples with a zero-shot prompt to create a few-shot prompt, which is subsequently provided to LLMs for equivalent mutant detection.
    Specifically, Figure\mbox{~\ref{fig:fsp_example}} illustrates the few-shot prompt template, which includes the same instructions as the zero-shot version along with the additional examples. 
    Each example consists of the code, the mutant, and the ground truth, accompanied by a rationale as an explanation. For our study, we manually wrote rationales for three randomly sampled examples from the training set.
    
    \item \textbf{Fine-tuning with instruction strategy}: it enables LLMs to gain domain-specific knowledge by training on numerous instruction-filled mutant pairs.
    We utilized the structured instruction format from our zero-shot prompting approach to create a fine-tuning dataset with the structure $<$\textit{mutant pair}, \textit{structured instruction}, \textit{equivalence}$>$. Subsequently, we fine-tuned the LLMs on this dataset to improve their ability to detect equivalent mutants using the zero-shot prompt.
\end{itemize}

\subsection{Baselines}
\label{subsec:baselines}
For a fair evaluation of LLM performance, we selected baseline techniques through a targeted literature review of SE publications from the past decade. We chose three well-established approaches that operate solely on code features without requiring test execution information and offer complete replication packages. These approaches provide ten distinct baselines for our comparative analysis:

\begin{itemize}[leftmargin=10pt]
    \item \textbf{Compiler-based technique}.
    \textit{Trivial Compiler Equivalence (TCE)} is an EMD technique that leverages compilation optimization~\cite{papadakis2015trivial,kintis2017detecting}. It works by using standard compilers to transform both the original program and its mutants into machine code, then determines mutant equivalence by comparing the resulting machine code.
    
    \item \textbf{ML-based technique}. 
    ~\citet{brito2020preliminary} extracted features from source code properties and control flow information, such as mutation operators and graph distances. Using these features, they developed seven machine learning classification models to identify equivalent mutants: \textit{K-Nearest Neighbors (KNN)}, \textit{Decision Tree (DT)}, \textit{Random Forest (RF)}, \textit{Support Vector Machine (SVM)}, \textit{Linear Discriminant Analysis (LDA)}, \textit{Logistic Regression (LR)}, and \textit{Gaussian Naive Bayes (GNB)}.
    
    \item \textbf{Tree-based neural network technique}. 
    ~\citet{peacock2021automatic} proposed the \textit{Abstract Syntax Tree Neural Network (ASTNN)} for equivalent mutant detection. 
    This approach uses ASTs of mutant programs as input and enhances detection performance by capturing both syntactical features (at lexical and statement levels) and code semantics through decomposing large ASTs and encoding multi-way statement trees.
\end{itemize}

We replicated baseline techniques according to implementations and parameter settings from previous research. 
As the original ML-based technique~\cite{brito2020preliminary} and Tree-based NN technique~\cite{peacock2021automatic} were limited to C code, we extended them to support Java code.
For comparison purposes, we included four variants of the widely-used TCE baseline: TCE$_{Javac}$ and TCE$_{Soot}$ for Java, and TCE$_{gcc}$ and TCE$_{clang}$ for C.

\subsection{Metrics}
\label{subsec:metrics}
\textbf{Effectiveness}.
In line with previous research~\cite{naeem2020machine,brito2020preliminary,chung2022augmenting,peacock2021automatic}, we evaluated our EMD techniques using standard binary classification metrics: \textit{Precision}, \textit{Recall}, and \textit{F1-score}.
The F1-score provides a balanced performance measure by calculating the harmonic mean of precision and recall.
To address potential class imbalance issues in our dataset, we employed macro-averaging for all metrics, calculating the unweighted mean across both classes. 
For instance, the macro-precision is computed as $ \frac{1}{2} \left( \frac{TP_0}{TP_0 + FP_0} + \frac{TP_1}{TP_1 + FP_1} \right) $, ensuring that the performance on the minority class is equally weighted and not overshadowed by the majority class. 
This provides a more realistic assessment of the technique's practical effectiveness.

Furthermore, to ensure our evaluation reflects the demanding conditions of real-world deployment, we implemented a strict penalty mechanism for compilation failures when evaluating compiler-based approaches. 
Due to inevitable misalignments between code versions and compiler versions, some code snippets and mutants fail to compile. 
Rather than excluding these incompilable pairs, which could artificially inflate results, we strictly penalized them as incorrect recommendations. 
For example, if a non-equivalent mutant (ground truth label 0) fails to compile, the recommendation is forced to 1. While this conservative penalty generates artificial false positives and decreases the reported precision, it guarantees a rigorous evaluation of the approaches under realistic constraints.

\textbf{Efficiency}. 
Detecting mutant pairs in mutation testing is time-intensive due to the substantial number of generated mutants. To quantify efficiency among equivalent mutant detection techniques, we compared their time overheads. We defined two metrics: (i) the average time to detect a mutant pair (referred to as \textit{inference time}), and (ii) the total time required to build an EMD model offline using the training set (referred to as \textit{training time}).

\subsection{Implementation and Environment}
\label{subsec:imp}
We obtained open-source pre-trained models (CodeBERT, GraphCodeBERT, PLBART, CodeT5, UniXCoder, CodeT5+, StarCoder, Code Llama, Llama 3, Qwen2.5-Coder, and Deepseek-Coder) from Huggingface~\cite{wolf2019huggingface}. We accessed Text-Embedding-Ada-002, Text-Embedding-3-Small, Text-Embedding-3-Large, GPT-3.5-Turbo (\texttt{gpt-3.5-turbo-0125}), and GPT-4 (\texttt{gpt-4-0613}) via OpenAI's APIs~\cite{openai2024}.
For our experiments, we adopted the recommended hyper-parameters from~\cite{lu2021codexglue} for both pre-trained and fine-tuned code embedding strategies to ensure effectiveness and fair comparison. Complete hyper-parameter settings are available on our project homepage~\cite{EMD2024}.
All experiments were performed on a machine with Intel Xeon CPU Gold-6342, 512 GB RAM, Ubuntu 20.04.6, and two A800 GPUs.
\section{Results}
\label{sec:results_and_analysis}

\subsection{RQ1: Performance of LLMs in EMD}
\label{subsec:RQ1}
\noindent
\textbf{\emph{\underline{Approach.}}}
This research question presents a 
analysis of various encoder LLMs’ performance on both Java and C scenarios, with particular emphasis on their code embedding capabilities.
Section~\ref{subsec:exp_plm} details the standard pre-trained code embedding methodology used by LLMs for equivalent mutant detection.
We trained all eight ML-based and Tree-based NN baselines using our training dataset for each programming language, following the identical settings and procedures outlined in their respective papers.
The Compiler-based baselines (TCE$_{Javac}$ and TCE$_{Soot}$ for Java, TCE$_{gcc}$ and TCE$_{clang}$ for C) leverage off-the-shelf compilers and required no training.
We evaluated the effectiveness of 10 state-of-the-art encoder LLMs against 10 baselines using precision, recall, and F1-score metrics.

\begin{table}[t]
\caption{The performance of baselines and state-of-the-art LLMs on equivalent mutant detection}
\label{tab:rq1}
\centering
\begin{tabular}{lrrrrrr} 
\toprule
\toprule
\multirow{2}{*}{\textbf{Technique}} & \multicolumn{3}{c}{\textbf{C}} & \multicolumn{3}{c}{\textbf{Java}} \\ \cmidrule(lr){2-4} \cmidrule(lr){5-7}
 & \textbf{Precision} & \textbf{Recall} & \textbf{F1-score} & \textbf{Precision} & \textbf{Recall} & \textbf{F1-score} \\ 
\midrule
\multicolumn{7}{l}{\framecolorbox[13.2cm][l]{gray!30}{gray!30}{\textbf{Compiler-based technique}}} \\ 
\midrule
TCE$_{gcc}$ & 5.93\% & 33.52\% & 10.08\% & -- & -- & --\\
TCE$_{clang}$ & 5.93\% & 33.52\% & 10.08\% & -- & -- & -- \\
TCE$_{Javac}$ & -- & -- & -- & 40.55\% & 38.15\% & 39.31\% \\
 TCE$_{Soot}$ & -- & -- & -- & 51.26\% & 51.81\% & 50.80\% \\
\midrule
\multicolumn{7}{l}{\framecolorbox[13.2cm][l]{gray!30}{gray!30}{\textbf{ML-based technique}}} \\
\midrule
KNN & 83.07\% & 65.05\% & 69.01\% & 77.77\% & 69.09\% & 72.15\% \\
DT & 70.24\% & 68.22\% & 69.12\% & 78.20\% & 67.20\% & 70.63\% \\
RF & 69.28\% & 84.00\% & 68.30\% & 78.89\% & 66.94\% & 70.53\% \\
SVM & 59.86\% & 55.15\% & 55.74\% & 93.62\% & 58.84\% & 61.61\% \\
LDA & 67.01\% & 52.63\% & 51.21\% & 88.02\% & 59.22\% & 62.09\% \\
LR & 65.46\% & 77.71\% & 60.38\% & 90.69\% & 59.33\% & 62.30\% \\
GNB & 65.01\% & 76.83\% & 59.16\% & 70.23\% & 62.10\% & 64.43\% \\ 
\midrule
\multicolumn{7}{l}{\framecolorbox[13.2cm][l]{gray!30}{gray!30}{\textbf{Tree-based NN technique}}} \\ 
\midrule
ASTNN & 41.38\% & 48.23\% & 44.55\% & 88.34\% & 65.27\% & 70.00\% \\ 
\midrule
\multicolumn{7}{l}{\framecolorbox[13.2cm][l]{gray!30}{gray!30}{\textbf{LLM-based technique (Pre-trained code embedding strategy)}}} \\ 
\midrule
CodeBERT (110M) & 41.64\% & 50.00\% & 45.44\% & 94.36\% & 64.26\% & 69.20\% \\
GraphCodeBERT (110M) & 75.25\% & 67.79\% & 70.39\% & 95.81\% & 74.30\% & 80.52\% \\
PLBART (210M) & 76.01\% & 74.38\% & \textbf{75.15\%} & 95.81\% & 74.30\% & 80.52\% \\
CodeT5 (210M) & \textbf{76.58\%} & 69.00\% & 71.70\% & 95.81\% & 74.30\% & 80.52\% \\
UniXCoder (110M) & 76.01\% & 74.38\% & \textbf{75.15\%} & 96.02\% & 74.30\% & 80.52\% \\
CodeT5+ (6B) & 75.55\% & 68.78\% & 71.26\% & \textbf{96.02\%} & \textbf{75.70\%} & \textbf{81.88\%} \\
StarCoder (7B) & 76.01\% & \textbf{74.38\%} & \textbf{75.15\%} & \textbf{95.99\%} & \textbf{75.70\%} & \textbf{81.88\%} \\
Text-Embedding-Ada-002 & 41.63\% & 50.00\% & 45.44\% & 94.99\% & 68.67\% & 74.56\% \\
Text-Embedding-3-Small & 60.21\% & 71.37\% & 62.43\% & 95.31\% & 70.88\% & 77.00\% \\
Text-Embedding-3-Large & 75.66\% & 68.34\% & 70.94\% & 95.96\% & 75.30\% & 81.50\% \\
\bottomrule
\bottomrule
\end{tabular}
\end{table}

\noindent
\textbf{\emph{\underline{Results.}}}
Table~\ref{tab:rq1} presents comparative results of LLMs against baseline approaches, evaluating precision, recall, and F1-score metrics across both Java and C programming environments.
First, in Java environment, our evaluation reveals that nearly all LLM-based techniques (with the exception of CodeBERT) yield higher F1-scores compared to baseline methods when measured by F1-score. 
For instance, the most effective LLM-based approaches (i.e., UniXCoder and CodeT5+) achieve an impressive F1-score of 81.88\%, higher than ASTNN (70.00\%), KNN (72.15\%), and TCE$_{Soot}$ (50.80\%).
On average, LLM-based techniques show F1-score increases of 75.18\%, 19.14\%, and 12.75\% over Compiler-based, ML-based, and Tree-based NN techniques, respectively.
The precision increases are 108.27\%, 15.90\%, and 8.23\%, while recall rates show differences of 62.05\%, 15.25\%, and 11.68\%, respectively.
Similar patterns emerge in the EMD for C scenario, where LLM-based techniques (except CodeBERT, Text-Embedding-Ada-002, and Text-Embedding-3-Small) also achieve higher F1-scores compared to all baselines. 
The highest-scoring LLM-based techniques (i.e., PLBART, UniXCoder, and StarCoder) achieve an F1-score of 75.15\%, compared to DT (69.12\%), ASTNN (44.55\%), and TCE$_{gcc}$ (10.08\%).
On average, in the C environment, LLM-based techniques show F1-score increases of 557.79\%, 7.21\%, and 48.83\% relative to Compiler-based, ML-based, and Tree-based NN techniques.
These results indicate the effectiveness of LLMs for equivalent mutant detection across both Java and C scenarios. This performance difference likely stems from LLMs being pre-trained on extensive code repositories, enabling them to develop a more nuanced understanding of code semantics compared to general ML/DL models that lack such specialized pre-training. This foundation makes LLMs particularly well-suited for equivalent mutant detection.

Second, our findings reveal that state-of-the-art Text-Embedding models underperform compared to pre-trained encoder LLMs in both Java and C scenarios. 
Notably, pre-trained encoder LLMs with fewer parameters (e.g., UniXCoder) yield higher F1-scores in equivalent mutant detection. 
For example, Text-Embedding-Ada-002 achieves F1-scores of 74.56\% and 45.44\% for Java and C, respectively, while UniXCoder reaches relatively higher F1-scores of 81.88\% and 75.15\%.
This performance gap likely stems from the fact that embedding models are general-purpose tools trained primarily on natural language data corpus. When applied to code-specific tasks like equivalent mutant detection, they suffer from the data-shift problem~\cite{ma2023scope}.
Our results suggest that smaller, code-specialized encoder LLMs generate code representations that better enable MLP-based classifiers to discern subtle semantic differences in code.

Third, different deployment scenarios place different emphasis on precision and recall. In settings where false positives are particularly costly, such as engineer-in-the-loop warning systems where each flagged mutant requires expensive manual inspection, precision becomes the most important metric. In the C dataset, the ML-based KNN model achieves the highest precision (83.07\%), outperforming LLM-based approaches such as CodeT5 (76.58\%), making it well suited for precision-critical applications. In contrast, for the Java dataset, LLM-based techniques demonstrate particularly strong reliability, with CodeT5+ and UniXCoder both achieving precision scores of 96.02\%, indicating a very low false-positive rate.
Conversely, in scenarios where the primary objective is to identify as many equivalent mutants as possible and minimize false negatives, recall becomes the key consideration. In the C dataset, traditional ML techniques again exhibit a clear advantage, with RF achieving the highest recall (84.00\%), substantially exceeding the best-performing LLM, StarCoder (74.38\%). This result highlights the classic precision–recall trade-off, as RF's superior recall is accompanied by a lower precision of 69.28\%. For the Java dataset, CodeT5+ and StarCoder achieve the highest recall (75.70\%), suggesting that LLM-based approaches can simultaneously maintain strong precision and competitive recall in this setting.

\begin{tcolorbox}\textbf{RQ1 Summary:}
Across both Java and C, we observed consistent performance patterns among the different EMD technique categories. 
LLMs achieved higher average F1-scores than all ten EMD baseline methods in equivalent mutant detection. 
In the Java scenario, LLMs showed F1-score increases of 75.18\% over Compiler-based techniques, 19.14\% over ML-based techniques, and 12.75\% over Tree-based NN techniques. 
Similarly, in the C context, LLMs showed increases of 557.79\% over Compiler-based approaches, 7.21\% over ML-based methods, and 48.83\% over Tree-based NN techniques in F1-score.
\end{tcolorbox}

\subsection{RQ2: Best strategy of LLMs in EMD}
\label{subsec:RQ2}
\noindent
\textbf{\emph{\underline{Approach.}}}
This research question investigates the impact of four additional LLM strategies (designed in Section~\ref{subsec:exp_llm}) on enhancing equivalent mutant detection compared to the pre-trained code embedding approach in RQ1.
For the fine-tuned code embedding strategy, we employed seven base models: CodeBERT, GraphCodeBERT, PLBART, CodeT5, UniXCoder, CodeT5+, and StarCoder.
For LLM-based approaches, we utilized Code Llama, Llama 3, Qwen2.5-Coder, Deepseek-Coder, GPT-3.5-Turbo, and GPT-4 to implement zero-shot prompting, few-shot prompting, and fine-tuning with instruction strategies. Following prior work~\cite{ma2023scope}, we adopted a 3-shot setting for few-shot prompting due to LLM input length constraints. Note that GPT-4 was excluded from the fine-tuning with instruction strategy due to API limitations.
We evaluated the effectiveness of these four LLM strategies by measuring precision, recall, and F1-score metrics.

\begin{table}
\caption{The performance of different LLM strategies on equivalent mutant detection}
\centering
\label{tab:rq2}
\begin{tabular}{lrrrrrr} 
\toprule\toprule
\multirow{2}{*}{\textbf{Technique}} & \multicolumn{3}{c}{\textbf{C}} & \multicolumn{3}{c}{\textbf{Java}} \\ \cmidrule(lr){2-4} \cmidrule(lr){5-7}
 & \textbf{Precision} & \textbf{Recall} & \textbf{F1-score} & \textbf{Precision} & \textbf{Recall} & \textbf{F1-score} \\ 
\midrule
\multicolumn{7}{l}{\framecolorbox[13.cm][l]{gray!30}{gray!30}{\textbf{Fine-tuned code embedding strategy}}} \\ 
\midrule
CodeBERT (110M) & 81.18\% & 75.70\% & 77.99\% & 90.39\% & 79.74\% & 83.87\% \\
GraphCodeBERT (110M) & 87.54\% & 90.43\% & 88.89\% & 91.54\% & 81.05\% & 85.18\% \\
PLBART (210M) & 89.78\% & \textbf{93.51\%} & 91.50\% & 93.24\% & 80.70\% & 85.42\% \\
CodeT5 (210M) & 87.25\% & 87.90\% & 87.57\% & 90.59\% & 80.34\% & 84.37\% \\
UniXCoder (110M) & 85.90\% & 91.85\% & 88.45\% & 94.33\% & 81.81\% & \textbf{86.58\%} \\
CodeT5+ (6B) & \textbf{90.88\%} & 92.30\% & \textbf{91.57\%} & 89.28\% & \textbf{82.79\%} & 85.59\% \\
StarCoder (7B) & 77.18\% & 72.40\% & 74.38\% & \textbf{96.02\%} & 75.70\% & 81.88\% \\ 
\midrule
\multicolumn{7}{l}{\framecolorbox[13.cm][l]{gray!30}{gray!30}{\textbf{Zero-shot prompting strategy}}} \\ 
\midrule
Code Llama (7B) & 58.62\% & 51.77\% & 18.12\% & 59.22\% & 50.78\% & 48.04\% \\
Llama 3 (8B) & 58.75\% & 52.65\% & 19.93\% & 50.60\% & 50.30\% & 49.49\% \\
Qwen2.5-Coder (7B) & 56.77\% & 55.97\% & 29.56\% & 60.15\% & 61.00\% & 60.53\% \\
Deepseek-Coder (6.7B) & 58.68\% & 52.21\% & 19.03\% & 57.14\% & 50.80\% & 48.29\% \\
GPT-3.5-Turbo & 53.12\% & 53.23\% & 30.14\% & 59.22\% & 59.70\% & 59.44\% \\
GPT-4 & 54.57\% & 51.44\% & 19.41\% & 67.42\% & 53.76\% & 53.61\% \\ 
\midrule
\multicolumn{7}{l}{\framecolorbox[13.cm][l]{gray!30}{gray!30}{\textbf{Few-shot prompting strategy}}} \\ 
\midrule
Code Llama (7B) & 58.47\% & 50.77\% & 16.01\% & 52.85\% & 50.38\% & 47.76\% \\
Llama 3 (8B) & 62.42\% & 72.11\% & 59.51\% & 46.31\% & 46.56\% & 46.43\% \\
Qwen2.5-Coder (7B) & 58.56\% & 56.29\% & 27.75\% & 56.77\% & 55.73\% & 56.12\% \\
Deepseek-Coder (6.7B) & 58.60\% & 51.66\% & 17.88\% & 61.26\% & 50.42\% & 47.03\% \\
GPT-3.5-Turbo & 54.36\% & 51.88\% & 21.22\% & 57.04\% & 52.23\% & 51.59\% \\
GPT-4 & 55.91\% & 53.54\% & 24.39\% & 67.02\% & 55.18\% & 55.90\% \\ 
\midrule
\multicolumn{7}{l}{\framecolorbox[13.cm][l]{gray!30}{gray!30}{\textbf{Fine-tuning with instruction strategy}}} \\ 
\midrule
Code Llama (7B) & 91.77\% & 56.39\% & 57.74\% & 93.21\% & 55.82\% & 56.79\% \\
Llama 3 (8B) & 73.45\% & 66.03\% & 68.51\% & 82.59\% & 72.40\% & 76.05\% \\
Qwen2.5-Coder (7B) & 64.23\% & 56.47\% & 57.50\% & 82.42\% & 70.53\% & 74.46\% \\
Deepseek-Coder (6.7B) & 73.94\% & 53.40\% & 52.40\% & 91.46\% & 74.47\% & 79.86\% \\
GPT-3.5-Turbo & 87.72\% & 89.99\% & 88.79\% & 92.82\% & 76.95\% & 82.31\% \\
\bottomrule\bottomrule
\end{tabular}
\end{table}

\noindent
\textbf{\emph{\underline{Results.}}} 
Tables~\ref{tab:rq1} and ~\ref{tab:rq2} present comparative results of five different LLM strategies and their effectiveness in equivalent mutant detection.
First, in Java environment, the fine-tuned UniXCoder achieves the highest F1-scores compared to all other LLM and strategy combinations, with F1-score improvements ranging from 1.16\% to 84.10\%.
For the C environment, the fine-tuned CodeT5+ yields the highest evaluation metrics, showing F1-score improvements between 0.08\% and 471.96\% over all other approaches.
These findings strongly suggest that utilizing smaller, domain-specific LLMs with fine-tuned code embedding strategies represents the most effective evaluated approach for equivalent mutant detection tasks.

Second, the fine-tuned code embedding strategy consistently achieves higher F1-scores than the pre-trained approach across almost all studied LLMs in terms of F1-score. 
In the Java environment, fine-tuning yields improvements of 21.20\%, 5.79\%, 6.09\%, 4.78\%, 5.74\%, and 4.53\% for CodeBERT, GraphCodeBERT, PLBART, CodeT5, UniXCoder, and CodeT5+, respectively.
The improvements are even more substantial in the C environment, with gains of 71.63\%, 26.28\%, 21.76\%, 22.13\%, 17.70\%, and 28.50\% for the same LLMs. StarCoder represents the only exception, showing no improvement and a slight decrease in F1-score in both Java and C environments.
These results clearly demonstrate that fine-tuned code embedding strategies can significantly enhance LLM performance in equivalent mutant detection. 
However, our analysis of recommendation results revealed an important trade-off: despite the overall higher performance of fine-tuned models, a small portion of pre-training knowledge is lost during fine-tuning. Specifically, an average of 1.09\% of the initially correct recommendations became incorrect after fine-tuning. This phenomenon highlights the ``catastrophic forgetting problem''~\cite{huang2023empirical} that can occur during the fine-tuning process of LLMs.

Third, code embedding strategies (pre-trained and fine-tuned embedding) consistently yield higher evaluation metrics than prompting strategies (zero-shot and few-shot prompting) across all evaluation metrics. In the Java environment, code embedding strategies demonstrate increased effectiveness with improvements of 37.29\%, 31.46\%, and 39.70\% in precision, recall, and F1-score, respectively. 
Similarly, in the C environment, the advantages are even more pronounced, with improvements of 24.63\%, 38.38\%, and a remarkable 127.60\% in F1-score.
This performance gap can be attributed to the fundamental difference in approach, prompting strategies require LLMs to comprehensively understand complex mutant pairs through context analysis, whereas code embedding strategies simply compare embedding vectors directly. 
The latter approach provides a more straightforward method for mutant understanding and comparison, resulting in the higher observed detection performance.

Fourth, the instruction-based fine-tuning strategy achieves higher scores across all metrics compared to both zero-shot and few-shot prompting approaches across all decoder-only LLMs on all metrics. 
In the Java environment, instruction-based fine-tuning achieved precision improvements of 48.04\% and 52.54\% over zero-shot and few-shot prompting, respectively, alongside recall improvements of 25.78\% and 31.55\%, and F1-score gains of 35.89\% and 41.94\%.
Similarly, in the C environment, instruction-based fine-tuning demonstrated precision improvements of 38.56\% and 36.55\% over zero-shot and few-shot approaches, with recall increases of 19.85\% and 20.07\%, and remarkable F1-score enhancements of 172.04\% and 194.90\%, respectively.
These results further demonstrate that fine-tuning strategies can significantly enhance LLM performance in equivalent mutant detection tasks.

Fifth, for engineering scenarios prioritizing the minimization of false positives (i.e., requiring high precision to avoid wasting time on false alarms), the fine-tuned code embedding strategy utilizing StarCoder achieves the highest precision in the Java environment (96.02\%).
In the C dataset, fine-tuning with instruction using Code Llama demonstrates a notable advantage, achieving a highly reliable precision of 91.77\%, making it particularly suitable for strict automated filtering.
Conversely, for exploratory tasks where avoiding false negatives is the priority (i.e., maximizing recall to capture all potential equivalent mutants), the fine-tuned code embedding strategy again shows specific utility depending on the underlying model. 
In the C dataset, PLBART excels with the highest recall of 93.51\%, while CodeT5+ leads in the Java environment with a recall of 82.79\%. 
These variances emphasize that rather than relying on a single dominant strategy, practitioners should select specific combinations of techniques and models that best align with their project's tolerance for false alarms versus missed equivalent mutants.

\begin{tcolorbox}\textbf{RQ2 Summary:}
Our experiments reveal performance differences across language environments. 
In the Java environment, fine-tuned UniXCoder demonstrates the highest F1-scores, with improvements ranging from 1.16\% to 84.10\% compared to all other LLM combinations and strategies. 
Similarly, for the C environment, fine-tuned CodeT5+ achieves higher metrics than alternatives, with F1-score improvements between 0.08\% and 471.96\%. 
These results clearly establish fine-tuned code embedding as the most effective evaluated strategy for equivalent mutant detection. 
Notably, LLMs using only prompting strategies yield significantly lower metrics.
\end{tcolorbox}

\subsection{RQ3: Orthogonality between Studied EMD Techniques}
\label{subsec:RQ3}
\noindent
\textbf{\emph{\underline{Approach.}}} 
This research question investigates the performance characteristics of various EMD techniques and LLM strategies across Java and C environments. 
To thoroughly examine their orthogonality, we conducted a comprehensive analysis from two distinct perspectives, building upon the experimental frameworks established in RQ1 and RQ2:
\begin{itemize}[leftmargin=10pt]
    \item \textbf{Between EMD categories.} 
    Drawing upon our RQ1 findings, we identified the best-performing EMD techniques across four distinct categories: Compiler-based, ML-based, Tree-based NN, and LLM-based techniques.
    Specifically, in Java environment, we selected TCE$_{Soot}$, KNN, ASTNN, and fine-tuned UniXCoder, with each technique representing one of the four categories.
    For C environment, we selected TCE$_{clang}$, DT, ASTNN, and fine-tuned CodeT5+, similarly representing the four respective categories.
    
    \item \textbf{Between LLM strategies.} 
    Building on our RQ2 findings, we identified the most effective EMD techniques among five LLM approaches: pre-trained code embedding, fine-tuned code embedding, zero-shot prompting, few-shot prompting, and fine-tuning with instruction strategies.
    Specifically, in Java environment, the selected EMD techniques are pre-trained UniXCoder, fine-tuned UniXCoder, ``Qwen2.5-Coder + zero-shot prompting'', ``Qwen2.5-Coder + few-shot prompting'', and ``GPT-3.5-Turbo + fine-tuning with instruction'', each representing their respective LLM strategies. 
    For C environment, the selected EMD techniques are pre-trained UniXCoder, fine-tuned CodeT5+, ``GPT-3.5-Turbo + zero-shot prompting'', ``Llama 3 + few-shot prompting'', and ``GPT-3.5-Turbo + fine-tuning with instruction'', similarly representing diverse LLM strategies.
\end{itemize}

Building on these perspectives, we conducted a two-level analysis: (1) \textbf{unique correct/incorrect detections}, and (2) \textbf{detection performance across mutation operators}.
For the first level, we employed Venn diagrams for both Java and C environments to visualize the unique correct/incorrect detections across the various EMD techniques studied.
For the second level, we examined each EMD technique's performance across different mutation operators in both Java and C environments by disaggregating detection results by mutation operator.
From a practical use-case perspective, this granular analysis allows test engineers to configure hybrid pipelines. 
For instance, confidently automating the silent removal of mutants from high-confidence recommended operators, while reserving manual IDE warnings for more complex, error-prone mutation types.
Additionally, we applied the \textit{Kruskal-Wallis test}~\cite{Kruskal-wallis-test}, a non-parametric statistical method for comparing multiple independent groups, to determine statistical significance in detection performance differences among the EMD techniques for each mutation operator.

\begin{figure*}[t!]
    \centering
    \includegraphics[width=1.0\linewidth]{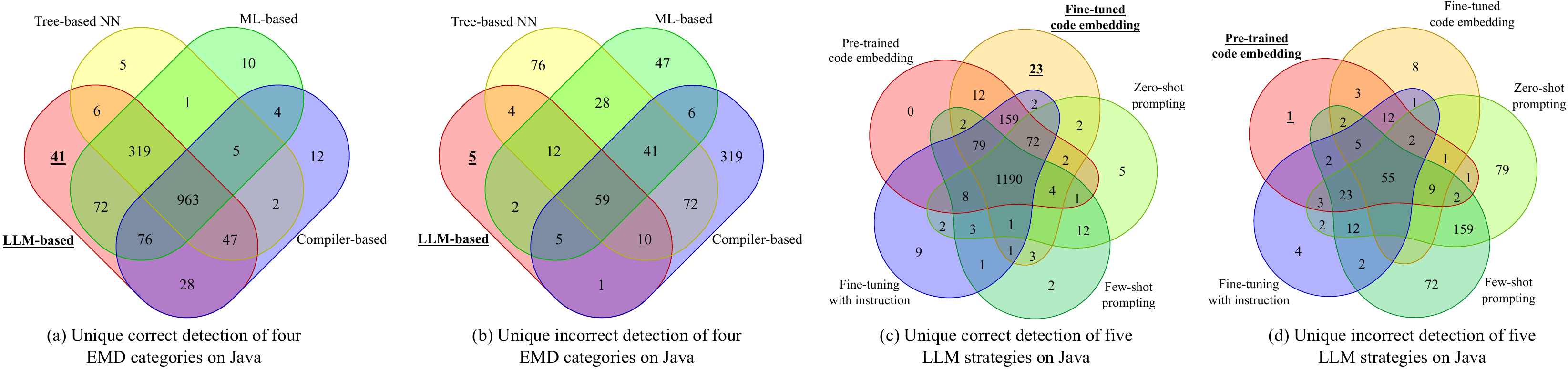}
    \caption{Unique correct detections ($\uparrow$) and unique incorrect detections ($\downarrow$) across studied techniques on Java}
    \label{fig:RQ3_veen}
\end{figure*}

\begin{figure*}[t!]
    \centering
    \includegraphics[width=1.0\linewidth]{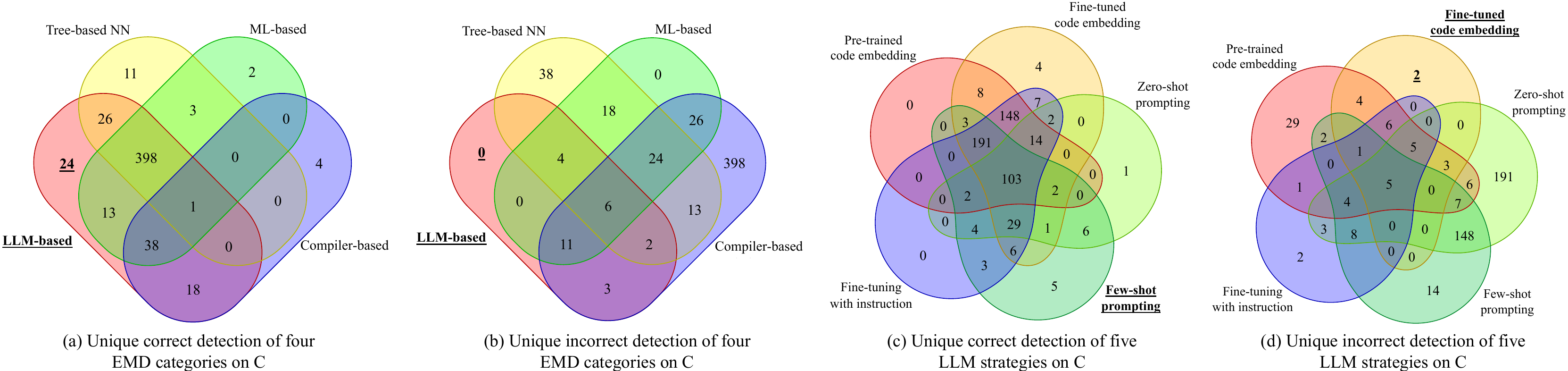}
    \caption{Unique correct detections ($\uparrow$) and unique incorrect detections ($\downarrow$) across studied techniques on C}
    \label{fig:RQ3_veen_c}
\end{figure*}

\begin{figure*}[t!]
    \centering
    \includegraphics[width=1.0\linewidth]{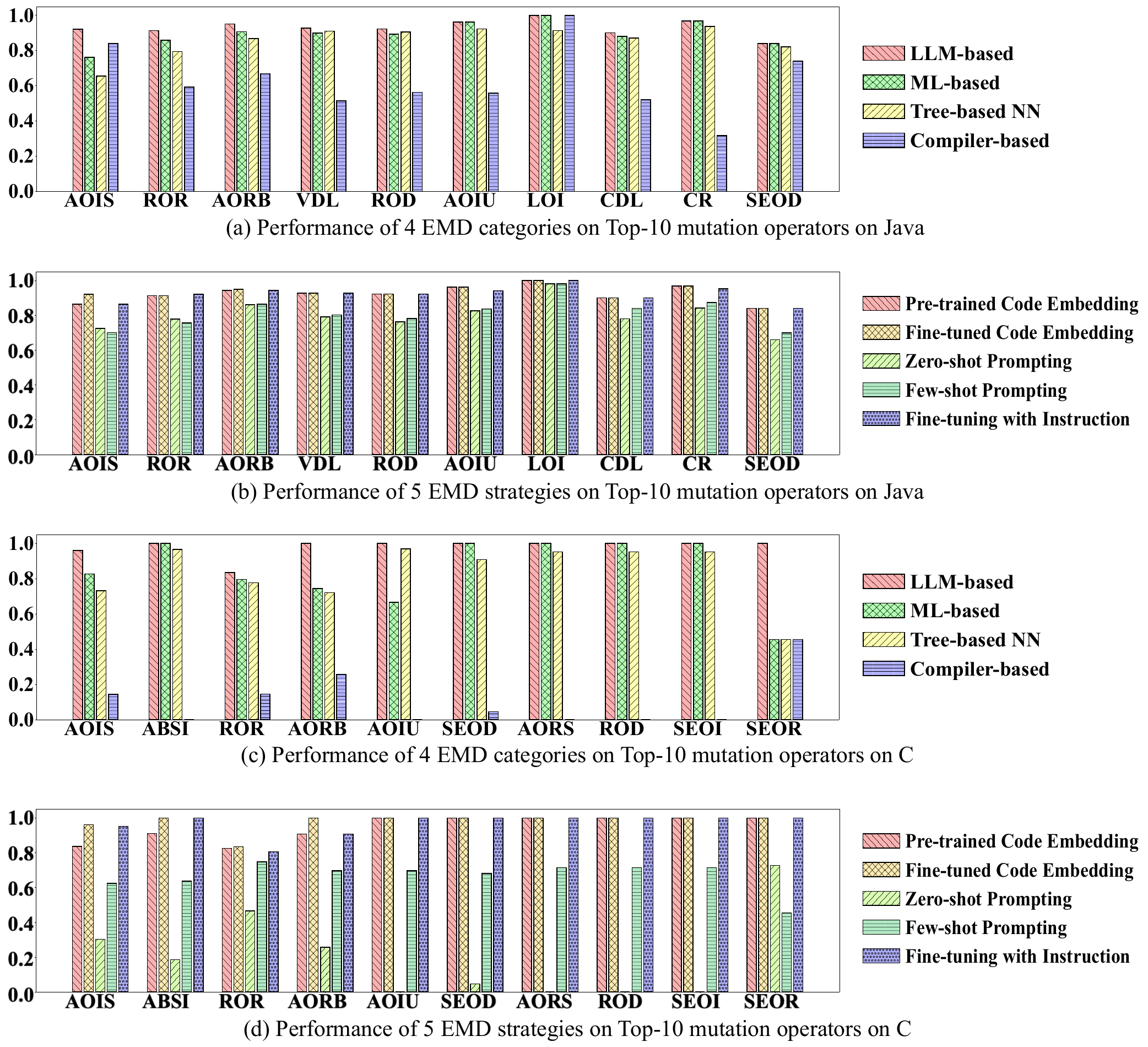}
    \caption{Detection performance on Top-10 mutation operators across various EMD techniques (x-axis shows mutation operators and y-axis shows the correct detection percentage)}
    \label{fig:RQ3barplot}
\end{figure*}

\noindent
\textbf{\emph{\underline{Results.}}} 
Figure~\ref{fig:RQ3_veen} and~\ref{fig:RQ3_veen_c} illustrate Venn diagrams showing the intersection of correct and incorrect detections across the studied EMD techniques from two analytical perspectives: EMD categories and LLM strategies. 
Overlapping areas indicate shared correct/incorrect detections among multiple techniques, while non-overlapping areas represent unique correct/incorrect detections specific to individual EMD techniques. 
Additionally, Figure~\ref{fig:RQ3barplot} displays the detection performance of each EMD technique across mutation operators. 
For brevity, we present results for only the top 10 most frequent mutation operators.

\textbf{Between EMD categories.}
Figure\mbox{~\ref{fig:RQ3_veen}(a)} demonstrates that the LLM-based technique identifies more unique \textit{correct} detections than all other EMD categories in the Java environment.
The LLM-based approach identified 41 unique correct detections, which is higher than the Compiler-based (12), ML-based (10), and Tree-based NN (5) techniques.
Furthermore, as shown in Figure~\ref{fig:RQ3_veen}(b), the LLM-based technique also excels by producing significantly fewer unique \textit{incorrect} detections. 
With only 5 unique incorrect detections, it demonstrates fewer false positives compared to Compiler-based (319), ML-based (47), and Tree-based NN (76) techniques.
These findings provide further quantitative support for the effectiveness of LLM-based approaches in equivalent mutant detection, corroborating our earlier results from RQ1.
Similarly, Figure~\ref{fig:RQ3_veen_c}(a) reveals that the LLM-based technique yields the highest number of unique \textit{correct} detections in the C environment. 
It successfully identified 24 unique correct detections, surpassing Compiler-based (4), ML-based (2), and Tree-based NN (11) approaches.
Figure~\ref{fig:RQ3_veen_c}(b) highlights the LLM-based technique's low false-positive rate, as it produced zero unique \textit{incorrect} detections. 
This demonstrates fewer \textit{incorrect} detections compared to Compiler-based (398) and Tree-based NN (38) techniques, while matching the precision of the ML-based approach (0).

Figure~\mbox{\ref{fig:RQ3barplot}}(a) demonstrates that the LLM-based technique consistently achieves higher detection rates than the other three EMD categories across nearly all mutation operators on Java environment.
The LLM-based technique achieves impressive \textit{correct} detection rates of 92.12\% (339), 91.28\% (314), and 95.03\% (287) for the three most prevalent mutation operators (\texttt{AOIS}, \texttt{ROR}, and \texttt{AROB}). In contrast, the second best performer, the ML-based technique, shows lower \textit{correct} detection rates of 76.09\% (280), 85.76\% (295), and 90.73\% (274), respectively.
The sole exception is the \texttt{COR} operator, where the LLM-based technique detects marginally fewer mutants than the best ML-based technique (13 vs. 14).
Similarly, Figure~\mbox{\ref{fig:RQ3barplot}}(c) illustrates that the LLM-based technique consistently achieves higher detection rates compared to the other three EMD categories across nearly all mutation operators in the C environment.
The LLM-based technique demonstrates superior performance with \textit{correct} detection rates of 96.10\% (222), 100.00\% (146), and 83.50\% (86) for the three most common mutation operators (\texttt{AOIS}, \texttt{ABSI}, and \texttt{ROR}). By comparison, the ML-based technique, which ranks second in effectiveness, achieves lower detection rates of 82.68\% (191), 100.00\% (146), and 79.61\% (82) for these same operators.
Statistical analysis using the \textit{Kruskal-Wallis test} confirms this difference with a significant \textit{p-value} of 8.89e-4 and 1.29e-3, establishing that the LLM-based technique statistically differs from all compared EMD categories in mutation operator detection in Java and C environments.

\textbf{Between LLM strategies.}
Figure~\mbox{\ref{fig:RQ3_veen}}(c) reveals that the fine-tuned code embedding strategy achieves the highest number of unique \textit{correct} detections in the Java environment.
Specifically, the fine-tuned code embedding strategy identifies 23 unique correct detections, exceeding the other approaches: 0 from pre-trained code embedding, 5 from zero-shot prompting, 2 from few-shot prompting, and 9 from fine-tuning with instruction.
Additionally, Figure~\ref{fig:RQ3_veen}(d) demonstrates that prompting-based strategies yield the highest number of unique \textit{incorrect} detections among all five approaches, with zero-shot and few-shot prompting producing 79 and 72 incorrect detections, respectively. 
In contrast, pre-trained code embedding generates only 1 incorrect detection, fine-tuned code embedding produces 8, and fine-tuning with instruction results in 4.
However, Figure~\mbox{\ref{fig:RQ3_veen_c}}(c) reveals that the few-shot prompting strategy identifies the most unique \textit{correct} detections in the C environment.
Specifically, the few-shot prompting strategy identifies 5 unique \textit{correct} detections, only slightly exceeding the fine-tuned code embedding with 4 unique \textit{correct} detections.
Additionally, Figure~\ref{fig:RQ3_veen_c}(d) demonstrates that the zero-shot prompting strategy yields the highest number of unique \textit{incorrect} detections among all five approaches, with 191 incorrect detections.
In contrast, fine-tuned code embedding generates only 2 \textit{incorrect} detections, pre-trained code embedding produces 29, fine-tuning with instruction results in 2, and few-shot prompting results in 14.
These findings indicate that LLMs relying solely on prompting strategies achieve lower metrics than other approaches for equivalent mutant detection, further corroborating our conclusions from RQ2.

Figure~\mbox{\ref{fig:RQ3barplot}}(b) demonstrates that the fine-tuned code embedding strategy consistently achieves higher detection rates than all other strategies across all mutation operators in the Java environment.
Notably, both zero-shot and few-shot prompting strategies consistently show the weakest detection performance for most mutation operators. For the three most common mutation operators (\texttt{AOIS}, \texttt{ROR}, and \texttt{AROB}), zero-shot prompting achieved correct detections of 267 (72.55\%), 268 (77.91\%), and 260 (86.09\%) respectively, while few-shot prompting reached 258 (70.10\%), 260 (75.58\%), and 261 (86.42\%). 
In contrast, the fine-tuned code embedding strategy yielded higher metrics with correct detections of 339 (92.12\%), 314 (91.28\%), and 287 (95.03\%), respectively.
Figure~\ref{fig:RQ3barplot}(d) also demonstrates that the fine-tuned code embedding strategy consistently achieves higher detection rates than all other strategies across all mutation operators in the C environment.
Similarly, both zero-shot and few-shot approaches consistently underperform for most mutation operators. For the three most prevalent mutation operators (\texttt{AOIS}, \texttt{ABSI}, and \texttt{ROR}), zero-shot prompting achieved modest detection rates of 30.30\% (70), 18.49\% (27), and 46.60\% (48), respectively. Few-shot prompting showed improvement with detection rates of 62.34\% (144), 63.70\% (93), and 74.76\% (77).
In contrast, the fine-tuned code embedding strategy demonstrated higher detection capabilities, achieving detection rates of 96.10\% (222), 100.00\% (146), and 83.50\% (86) for the same operators, a substantial improvement over both solely using prompting approaches.
Statistical analysis using the \textit{Kruskal-Wallis test} confirms this difference with a significant \textit{p-value} of 7.84e-09 and 7.12e-09,  providing strong evidence for the effectiveness of the fine-tuned code embedding approach in both Java and C environments.

\begin{tcolorbox}\textbf{RQ3 Summary:}
Our evaluation reveals that both the LLM-based technique and the fine-tuned code embedding approach achieve higher effectiveness than alternative EMD categories and LLM strategies. 
This capability is demonstrated through higher correct detection rates, lower incorrect detection rates, and consistent effectiveness across diverse mutation operators in both C and Java programming languages.
\end{tcolorbox}

\subsection{RQ4: Efficiency of Studied EMD Techniques}
\label{subsec:RQ4}
\noindent
\textbf{\emph{\underline{Approach.}}} 
We evaluate the efficiency of our studied EMD techniques by measuring both training time (total time required to build an EMD model offline) and inference time (average time needed to detect a mutant pair) on both Java and C environments.
Since the training phase is not applicable to all EMD techniques (such as TCE) and occurs only once offline before inference, we primarily use inference time as the key metric for comparing efficiency across techniques.
In a practical development workflow, this computationally intensive training phase is fully decoupled from daily operations and managed within a centralized model maintenance process.  
Retraining is only required periodically, such as during major version releases or when new mutation operators are introduced, ensuring that the fast inference times remain the sole factor impacting a developer's real-time IDE experience or the overall speed of automated testing pipelines.

\begin{table}
\caption{Time efficiency of studied EMD techniques}
\label{tab:rq4}
\centering
\begin{adjustbox}{max width=1.0 \textwidth,center}
\begin{tabular}{llrrrr} 
\toprule
\toprule
 \multicolumn{2}{c}{\multirow{3}{*}{\textbf{Technique}}} & \multicolumn{2}{c}{\textbf{C}} & \multicolumn{2}{c}{\textbf{Java}} \\ \cmidrule(lr){3-4} \cmidrule(lr){5-6}
 &  & \multicolumn{1}{c}{\begin{tabular}[c]{@{}c@{}}\textbf{Tranining }\\\textbf{Time (s)}\end{tabular}} & \multicolumn{1}{c}{\begin{tabular}[c]{@{}c@{}}\textbf{ Inference }\\\textbf{Time (s)}\end{tabular}} & \multicolumn{1}{c}{\begin{tabular}[c]{@{}c@{}}\textbf{Tranining }\\\textbf{Time (s)}\end{tabular}} & \multicolumn{1}{c}{\begin{tabular}[c]{@{}c@{}}\textbf{ Inference }\\\textbf{Time (s)}\end{tabular}} \\ 
\midrule
\multirow{2}{*}{\textbf{Compiler-based}} & TCE$_{gcc}$ / TCE$_{Javac}$ & - & 1.5265 & - & 1.0241 \\
 & TCE$_{clang}$ / TCE$_{Soot}$ & - & 0.6613 & - & 2.3537 \\ 
\midrule
\multirow{7}{*}{\textbf{ML-based}} & KNN & 125.4208 & 0.0024 & 298.8415 & 0.0019 \\
 & DT & 125.5198 & 0.0026 & 297.3026 & 0.0015 \\
 & RF & 126.1018 & 0.0037 & 300.3978 & 0.0081 \\
 & SVM & 125.5248 & 0.0026 & 297.4997 & 0.0018 \\
 & LDA & 125.2258 & 0.0021 & 297.7096 & 0.0016 \\
 & LR & 125.4288 & 0.0024 & 296.8087 & 0.0016 \\
 & GNB & 125.3298 & 0.0022 & 298.2195 & 0.0014 \\ 
\midrule
\textbf{Tree-based NN} & ASTNN & 3895.6642 & 0.2592 & 306.7047 & 0.0274 \\ 
\midrule
\multicolumn{6}{l}{\framecolorbox[14.5cm][l]{gray!30}{gray!30}{\textbf{LLM-based Technique}}} \\ 
\midrule
\multirow{10}{*}{\begin{tabular}[c]{@{}l@{}}\textbf{Pre-trained }\\\textbf{Code Embedding}\end{tabular}} & CodeBERT (110M) & 180.3765 & 0.0287 & 562.616 & 0.0269 \\
 & GraphCodeBERT (110M) & 308.3287 & 0.03628 & 805.1435 & 0.0429 \\
 & PLBART (210M) & 291.7131 & 0.05009 & 844.1389 & 0.0421 \\
 & CodeT5 (210M) & 891.6265 & 0.0784 & 1545.3771 & 0.0784 \\
 & UniXCoder (110M) & 219.565 & 0.02151 & 809.1785 & 0.0431 \\
 & CodeT5+ (6B) & 6265.9362 & 0.57782 & 17043.0572 & 0.8294 \\
 & StarCoder (7B) & 7328.3786 & 0.56932 & 16634.3038 & 0.9292 \\
 & Text-Embedding-Ada-002 & 3523.3168 & 0.6468 & 9820.2909 & 0.5951 \\
 & Text-Embedding-3-Small & 3576.3768 & 0.6566 & 11346.9648 & 0.6876 \\
 & Text-Embedding-3-Large & 6165.2026 & 1.1329 & 19234.9228 & 1.1705 \\ 
\hdashline
\multirow{7}{*}{\begin{tabular}[c]{@{}l@{}}\textbf{Fine-tuned }\\\textbf{Code Embedding}\end{tabular}} & CodeBERT (110M) & 599.3331 & 0.02531 & 1734.3351 & 0.0269 \\
 & GraphCodeBERT (110M) & 552.2314 & 0.03219 & 2613.7416 & 0.0429 \\
 & PLBART (210M) & 954.9454 & 0.04952 & 2390.2443 & 0.0421 \\
 & CodeT5 (210M) & 978.2492 & 0.04781 & 4471.2962 & 0.0784 \\
 & UniXCoder (110M) & 873.2314 & 0.01392 & 2566.1184 & 0.0431 \\
 & CodeT5+ (6B) & 8684.8573 & 0.72498 & 37286.3283 & 0.8294 \\
 & StarCoder (7B) & 13693.9018 & 0.6518 & 41888.536 & 0.9292 \\ 
\hdashline
\multirow{5}{*}{\begin{tabular}[c]{@{}l@{}}\textbf{Zero-shot }\\\textbf{prompting}\end{tabular}} & Code Llama (7B) & - & 0.1933 & - & 0.2068 \\
 & Qwen2.5 (7B) & - & 0.1239 & - & 0.1312 \\
 & Deepseek-Coder (6.7B) & - & 0.1529 & - & 0.1509 \\
 & GPT-3.5-Turbo & - & 0.4607 & - & 0.499 \\
 & GPT-4 & - & 0.806 & - & 0.5808 \\ 
\hdashline
\multirow{5}{*}{\begin{tabular}[c]{@{}l@{}}\textbf{Few-shot }\\\textbf{prompting}\end{tabular}} & Code Llama (7B) & - & 0.4154 & - & 0.5639 \\
 & Qwen2.5 (7B) & - & 0.3015 & - & 0.4076 \\
 & Deepseek-Coder (6.7B) & - & 0.3671 & - & 0.4996 \\
 & GPT-3.5-Turbo & - & 0.5219 & - & 0.529 \\
 & GPT-4 & - & 0.8951 & - & 0.6601 \\ 
\hdashline
\multirow{5}{*}{\begin{tabular}[c]{@{}l@{}}\textbf{Fine-tuning }\\\textbf{with Instruction}\end{tabular}} & Code Llama (7B) & 9613.5899 & 0.4648 & 29206.5457 & 0.5286 \\
 & Llama 3 (8B) & 6845.9096 & 0.3139 & 23076.1362 & 0.3283 \\
 & Qwen2.5-Coder  (7B) & 6755.4977 & 0.2178 & 22007.7334 & 0.2374 \\
 & Deepseek-Coder (6.7B) & 8702.1011 & 0.2293 & 26672.1323 & 0.23 \\
 & GPT-3.5-Turbo & 1702.0035 & 0.6817 & 6976.0079 & 0.3156 \\
\bottomrule
\bottomrule
\end{tabular}
\end{adjustbox}
\end{table}

\noindent
\textbf{\emph{\underline{Results.}}}
From Table~\ref{tab:rq4}, in Java environment, we observe that the inference times for the best-performing techniques across different categories are Compiler-based (TCE$_{Soot}$) at \SI{2.3537}{s}, ML-based (KNN) at \SI{0.0019}{s}, Tree-based NN (ASTNN) at \SI{0.0274}{s}, and LLM-based (UniXCoder) at \SI{0.0431}{s}.
In C environment, the inference times for the best-performing techniques are Compiler-based (TCE$_{clang}$) at \SI{0.6613}{s}, ML-based (LDA) at \SI{0.0021}{s}, Tree-based NN (ASTNN) at \SI{0.2592}{s}, and LLM-based (UniXCoder) at \SI{0.0215}{s}.
LLM-based techniques require less inference time and yield higher F1-scores when compared to traditional Compiler-based techniques. 
Although LLM-based techniques may consume slightly more computational resources than ML-based and Tree-based NN techniques, their higher evaluation metrics present a measurable trade-off for this additional investment. 
Ultimately, LLM-based techniques provide an practical balance between computational cost and effectiveness, making them a viable option for EMD.

The pre-trained code embedding strategy requires significantly less training time than the fine-tuned approach in terms of training time. For instance, pre-trained UniXCoder requires only \SI{809.1785}{s} and \SI{219.565}{s} for Java and C environments, while fine-tuned UniXCoder needs \SI{2566.1184}{s} and \SI{873.2314}{s}, making the latter approximately three times slower. 
This demonstrates that limited computational resources combined with large LLM model sizes are important factors to consider when choosing a code embedding strategy to enhance LLM performance. 
Consequently, practical implementations should balance efficiency against resource consumption when deploying LLM-based techniques.

\begin{tcolorbox}\textbf{RQ4 Summary:}
The inference time of the best-performing LLM-based technique (\SI{0.0431}{s} for Java and \SI{0.0215}{s} for C) is faster than the best-performing Compiler-based technique (\SI{2.3537}{s} for Java and \SI{0.6613}{s} for C). However, it is slightly slower than both the best-performing ML-based technique (\SI{0.0019}{s} for Java and \SI{0.0021}{s} for C) and the best-performing Tree-based NN technique (\SI{0.0274}{s} for Java).
Given the significant effectiveness of LLM-based techniques, this minor increase in inference time is acceptable, demonstrating a practical trade-off between computational cost and performance effectiveness.
\end{tcolorbox}

\subsection{RQ5: Performance of Fine-tuned LLMs in Cross-lingual EMD}
\label{subsec:RQ5}
\noindent
\textbf{\emph{\underline{Approach.}}} 
Our results demonstrated the effectiveness of fine-tuned LLMs on EMD for Java and C when trained separately. 
However, real-world development often involves multiple programming languages, and training separate LLMs for each language is costly.
This research question examines the cross-lingual generalization ability of the studied LLMs by fine-tuning them on both Java and C simultaneously and then evaluating each language individually.
We combined the Java training dataset \textit{$MutantBench_{train-Java}$} and C training dataset \textit{$MutantBench_{train-C}$} into a cross-lingual training dataset \textit{$MutantBench_{cross-lang}$}. 
We then fine-tuned the LLMs on this cross-lingual dataset and evaluated their performance separately on the Java test dataset \textit{$MutantBench_{test-Java}$} and C test dataset \textit{$MutantBench_{test-C}$}.

Based on quantitative results, we conducted an operator-level analysis using radar diagrams to understand performance differences between LLMs fine-tuned on single-lingual datasets (e.g., C or Java) versus cross-lingual datasets for the Top-10 mutant operators. 
We selected the ten most common operators in the dataset as the Top-10 mutant operators.
Based on our two categories of LLM fine-tuning approaches (fine-tuned code embedding and fine-tuned with instruction), we selected the best-performing LLM for each category. This selection was made by calculating the weighted F1-score that combines results from both single-lingual and cross-lingual dataset training ($\frac{1}{2}*F1_{single-lang}+\frac{1}{2}*F1_{cross-lang}$).
For the fine-tuned code embedding strategy, we employed PLBART for C and GraphCodeBERT for Java, respectively. 
In the case of the fine-tuned with instruction strategy, we selected GPT-3.5-Turbo for both C and Java.

\begin{table}
\caption{The performance of different fine-tuned LLMs on cross-lingual equivalent mutant detection}
\centering
\label{tab:rq5}
\begin{tabular}{lrrrrrr}
\toprule
\toprule
\multirow{2}{*}{\textbf{Technique}} & \multicolumn{3}{c}{\textbf{C}} & \multicolumn{3}{c}{\textbf{Java}} \\ \cmidrule(lr){2-4} \cmidrule(lr){5-7}
 & \textbf{Precision} & \textbf{Recall} & \textbf{F1-score} & \textbf{Precision} & \textbf{Recall} & \textbf{F1-score} \\ 
\midrule
\multicolumn{7}{l}{\framecolorbox[13.2cm][l]{gray!30}{gray!30}{\textbf{Fine-tuned LLMs} (Training on $MutantBench_{train-C}$ / \textit{$MutantBench_{train-Java}$})}} \\ 
\midrule
CodeBERT (110M) & 81.18\% & 75.70\% & 77.99\% & 90.39\% & 79.74\% & 83.87\% \\
GraphCodeBERT (110M) & 87.54\% & 90.43\% & 88.89\% & 91.54\% & 81.05\% & 85.18\% \\
PLBART (210M) & 89.78\% & 93.51\% & 91.50\% & 93.24\% & 80.70\% & 85.42\% \\
CodeT5 (210M) & 87.25\% & 87.90\% & 87.57\% & 90.59\% & 80.34\% & 84.37\% \\
UniXCoder (110M) & 85.90\% & 91.85\% & 88.45\% & 94.33\% & 81.81\% & 86.58\% \\
CodeT5+ (6B) & 90.88\% & 92.30\% & 91.57\% & 89.28\% & 82.79\% & 85.59\% \\
StarCoder (7B) & 77.18\% & 72.40\% & 74.38\% & 96.02\% & 75.70\% & 81.88\% \\ 
\hdashline
Code Llama (7B) & 91.77\% & 56.39\% & 57.74\% & 93.21\% & 55.82\% & 56.79\% \\
Llama 3 (8B) & 73.45\% & 66.03\% & 68.51\% & 82.59\% & 72.40\% & 76.05\% \\
Qwen2.5-Coder (7B) & 64.23\% & 56.47\% & 57.50\% & 82.42\% & 70.53\% & 74.46\% \\
Deepseek-Coder (6.7B) & 73.94\% & 53.40\% & 52.40\% & 91.46\% & 74.47\% & 79.86\% \\
GPT-3.5-Turbo & 87.72\% & 89.99\% & 88.79\% & 92.82\% & 76.95\% & 82.31\% \\
\midrule
\multicolumn{7}{l}{\framecolorbox[13.2cm][l]{gray!30}{gray!30}{\textbf{Fine-tuned LLMs} (Training on \textit{$MutantBench_{cross-lang}$})}} \\  
\midrule
CodeBERT (110M) & 86.60\% & 87.24\% & 86.92\% & 94.57\% & 82.81\% & 87.39\% \\
GraphCodeBERT (110M) & 90.99\% & 94.83\% & 92.76\% & 93.50\% & 83.64\% & 87.64\% \\
PLBART (210M) & 91.14\% & 91.86\% & 91.50\% & 94.28\% & 77.70\% & 83.28\% \\
CodeT5 (210M) & 83.91\% & 83.61\% & 83.76\% & 91.43\% & 79.88\% & 84.28\% \\
UniXCoder (110M) & 88.61\% & 93.18\% & 90.67\% & 90.66\% & 80.54\% & 84.53\% \\
CodeT5+ (6B) & 75.55\% & 68.78\% & 71.26\% & 96.02\% & 75.70\% & 81.88\% \\
StarCoder (7B) & 70.69\% & 71.74\% & 69.98\% & 96.02\% & 75.70\% & 81.88\% \\
\hdashline
Code Llama (7B) & 77.06\% & 63.84\% & 67.02\% & 94.75\% & 74.99\% & 80.98\% \\
Llama 3 (8B) & 74.84\% & 67.24\% & 69.84\% & 90.70\% & 75.94\% & 80.96\% \\
Qwen2.5-Coder (7B) & 75.37\% & 72.84\% & 73.98\% & 83.60\% & 76.97\% & 79.72\% \\
Deepseek-Coder (6.7B) & 74.18\% & 65.26\% & 67.98\% & 94.77\% & 73.02\% & 79.10\% \\
GPT-3.5-Turbo & 85.94\% & 93.28\% & 88.96\% & 94.77\% & 81.28\% & 86.30\% \\
\bottomrule
\bottomrule
\end{tabular}
\end{table}

\noindent
\textbf{\emph{\underline{Results.}}}
Table~\ref{tab:rq5} compares the performance of LLMs fine-tuned on C and Java individually against those fine-tuned on cross-lingual data (i.e., \textit{$MutantBench_{cross-lang}$}) that incorporates both languages.
First, on both Java and C environments, results show that most fine-tuned LLMs on cross-lingual data achieve superior effectiveness compared to those individually fine-tuned on Java or C. 
On average, the fine-tuned LLMs on cross-lingual data achieve the precision of 92.92\%, recall of 78.18\%, and F1-score of 83.16\% in Java environment, and the precision of 81.24\%, recall of 79.48\%, and F1-score of 79.55\% in C environment.
While fine-tuned on cross-lingual data, the fine-tuned LLMs improve 2.50\% (on Java) and 3.23\% (on C) higher recall than those individually fine-tuned on Java or C, 3.70\% (on Java) and 3.55\% (on C) higher F1-score, respectively.
It demonstrates that the fine-tuned LLMs on cross-lingual data can enhance the performance of LLMs in equivalent mutant detection between syntactically similar languages.

Second, while training on cross-lingual data generally improves LLM performance in equivalent mutant detection, the degree of improvement varies across different LLM strategies, and in some cases, performance actually decreases.
For instance, LLMs with fine-tuned code embedding strategy trained on cross-lingual data (i.e., CodeBERT, GraphCodeBERT, PLBART, CodeT5, UniXCoder, CodeT5+, and StarCoder) achieve average precision of 93.78\% and 84.46\% on Java and C environments, respectively, with average recall of 76.44\% and 73.56\%, and average F1-scores of 81.41\% and 73.56\%.
In contrast, when trained on Java or C individually, these same LLMs achieve average precision of 92.20\% and 85.67\%, average recall of 80.30\% and 86.30\%, and average F1-scores of 84.70\% and 85.76\%, respectively.
However, LLMs using fine-tuning with instruction show significant performance improvements with cross-lingual fine-tuning.
Specifically, fine-tuning with instruction on cross-lingual data (i.e., Code Llama, Llama 3, Qwen2.5-Coder, Deepseek-Coder, GPT-3.5-Turbo) demonstrate higher precision than those individually fine-tuned on Java or C by 20.63\% (on Java) and 0.42\% (on C), higher recall by 9.15\% (on Java) and 13.36\% (on C), and higher F1-scores by 10.17\% (on Java) and 14.38\% (on C).
This comparison reveals that cross-lingual fine-tuning decreases the performance of fine-tuned code embedding strategy but boosts the performance of fine-tuning with instruction on equivalent mutant detection.

\begin{figure}[t!]
    \centering
    \includegraphics[width=0.9\linewidth]{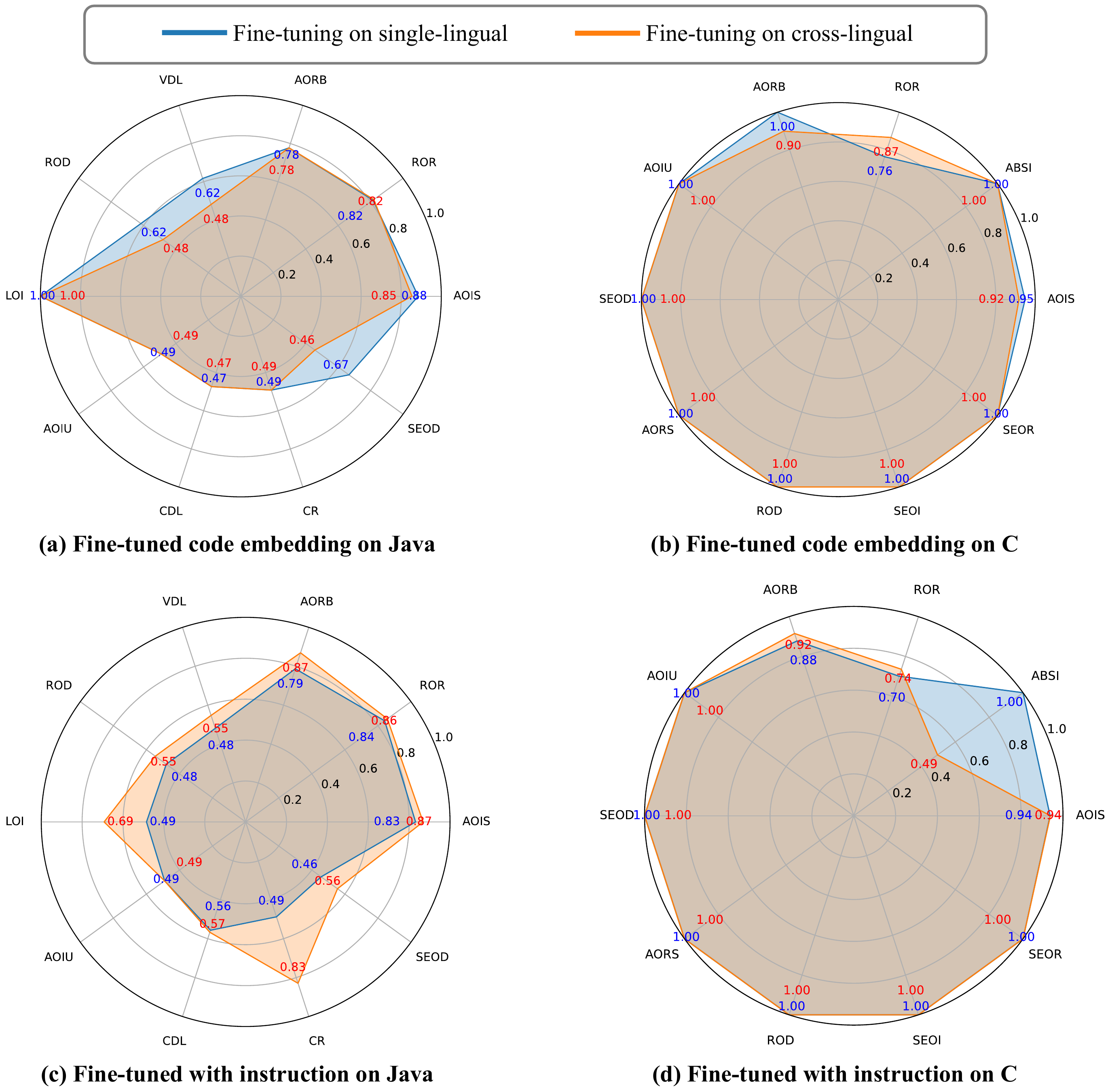}
    \caption{F1-score by Top-10 mutant operators}
    \label{fig:radar}
\end{figure}

Third, Figure~\ref{fig:radar}a shows that for the Top-10 Java mutant operators, the fine-tuned code embedding strategy on the single-lingual dataset achieves higher F1-scores across most of Top-10 Java mutant operators compared to the cross-lingual dataset. 
Specifically, the single-lingual approach yields F1-scores of 0.62 for \texttt{ROR}, 0.62 for \texttt{VDL}, 0.88 for \texttt{AOIS}, and 0.67 for \texttt{SEOD}. In contrast, the cross-lingual approach only reaches F1-scores of 0.48 for \texttt{ROR}, 0.48 for \texttt{VDL}, 0.85 for \texttt{AOIS}, and 0.46 for \texttt{SEOD}.
Figure~\ref{fig:radar}b demonstrates that the fine-tuned code embedding strategy on the single-lingual dataset achieves F1-scores equal to those of the cross-lingual dataset across most of the Top-10 C mutant operators (with only three exceptions).
For example, the single-lingual approach outperforms the cross-lingual approach on \texttt{AORB} and \texttt{AOIS} in terms of F1-score (1.00 vs. 0.90 and 0.95 vs. 0.92). However, the single-lingual approach performs worse than the cross-lingual approach on \texttt{ROR} (0.76 vs. 0.87).
These results indicate that fine-tuning on cross-lingual dataset may bring more negative benefits to the fine-tuned code embedding strategy.

Fourth, Figure~\ref{fig:radar}c shows that for the Top-10 Java mutant operators, the fine-tuned with instruction strategy on the single-lingual dataset achieves higher F1-scores across most mutant operators (with only two exceptions).
For instance, the cross-lingual approach yields F1-scores of 0.83 for \texttt{CR} and 0.69 for \texttt{LOI}. In contrast, the single-lingual approach only reaches F1-scores of 0.49 for both \texttt{CR} and \texttt{LOI}.
Figure~\ref{fig:radar}d shows that the fine-tuned with instruction strategy on the single-lingual dataset achieves F1-scores equal to those of the cross-lingual dataset across most of the Top-10 C mutant operators (with only three exceptions).
For instance, the single-lingual approach significantly outperforms the cross-lingual approach on \texttt{ABSI} (1.00 vs. 0.49). However, the cross-lingual approach outperforms the single-lingual approach on \texttt{AORB} and \texttt{ROR} (0.92 vs. 0.88 and 0.74 vs. 0.70).
Based on the quantitative results and operator-level analysis, fine-tuning on the cross-lingual dataset tends to improve the effectiveness when fine-tuned with the instruction strategy for structurally similar languages. 
However, we note that both C and Java are strongly typed and share a curly-brace syntax. 
The extent to which these findings generalize to dynamically typed languages (e.g., JavaScript) or languages with substantially different syntactic paradigms (e.g., Python or Visual Basic) remains an open question and represents an important direction for future research.

\begin{tcolorbox}\textbf{RQ5 Summary:}
Fine-tuning on cross-lingual data can improve the effectiveness of fine-tuned LLMs on equivalent mutant detection with both Java and C.
Overall, LLMs fine-tuned on cross-lingual data show superior performance, achieving higher F1-score on both Java and C environments. 
On average, cross-lingual fine-tuning leads to a 3.55\%$\sim$3.70\% improvement in F1-score over single-language fine-tuning. 
However, the effectiveness of cross-lingual fine-tuning varies based on LLM strategies. 
Fine-tuned code embedding strategy tend to perform worse with cross-lingual fine-tuning, showing decreases in recall and F1-score. 
In contrast, fine-tuning with instructions benefits substantially, with significant improvements across all metrics, especially in recall and F1-score.
\end{tcolorbox}

\section{Discussion}
\label{sec:discussion}

\subsection{Lessons Learnt}
\subsubsection{Model Architecture and Pre-training Diversity}
Our study empirically validated the performance of various LLMs to understand how the diversity in model architectures and pre-training paradigms impacts equivalent mutant detection. 
Initially, we assumed larger LLMs would have broader prior knowledge and increased learning capacity, thus improving performance. 
However, our experiments revealed that mere model size is the least relevant factor influencing LLM performance in this task.

This pattern appears in both C and Java environments, where the largest LLMs typically do not achieve the best results. 
Instead, our findings indicate that the diversity in data modality and pre-training tasks are the most influential factors, which aligns with conclusions from previous research\mbox{\cite{guo2022unixcoder,wang2023codet5+,gao-etal-2021-simcse}}.

For example, UniXCoder outperforms all other studied LLMs in the Java environment, as shown in RQ1 and RQ2. 
This superiority likely stems from UniXCoder's specific architectural design, which incorporates Abstract Syntax Trees (AST) to enhance code embeddings with rich syntax and semantic information. 
Additionally, both UniXCoder and CodeT5+ employ contrastive learning with multiple well-designed code-related pre-training tasks. 
Ultimately, these findings demonstrate that the rich diversity in how models are structured and aligned to code semantics provides much deeper lessons for equivalent mutant detection than scaling model parameters.

\begin{figure}[t!]
    \centering
    \includegraphics[width=1\linewidth]{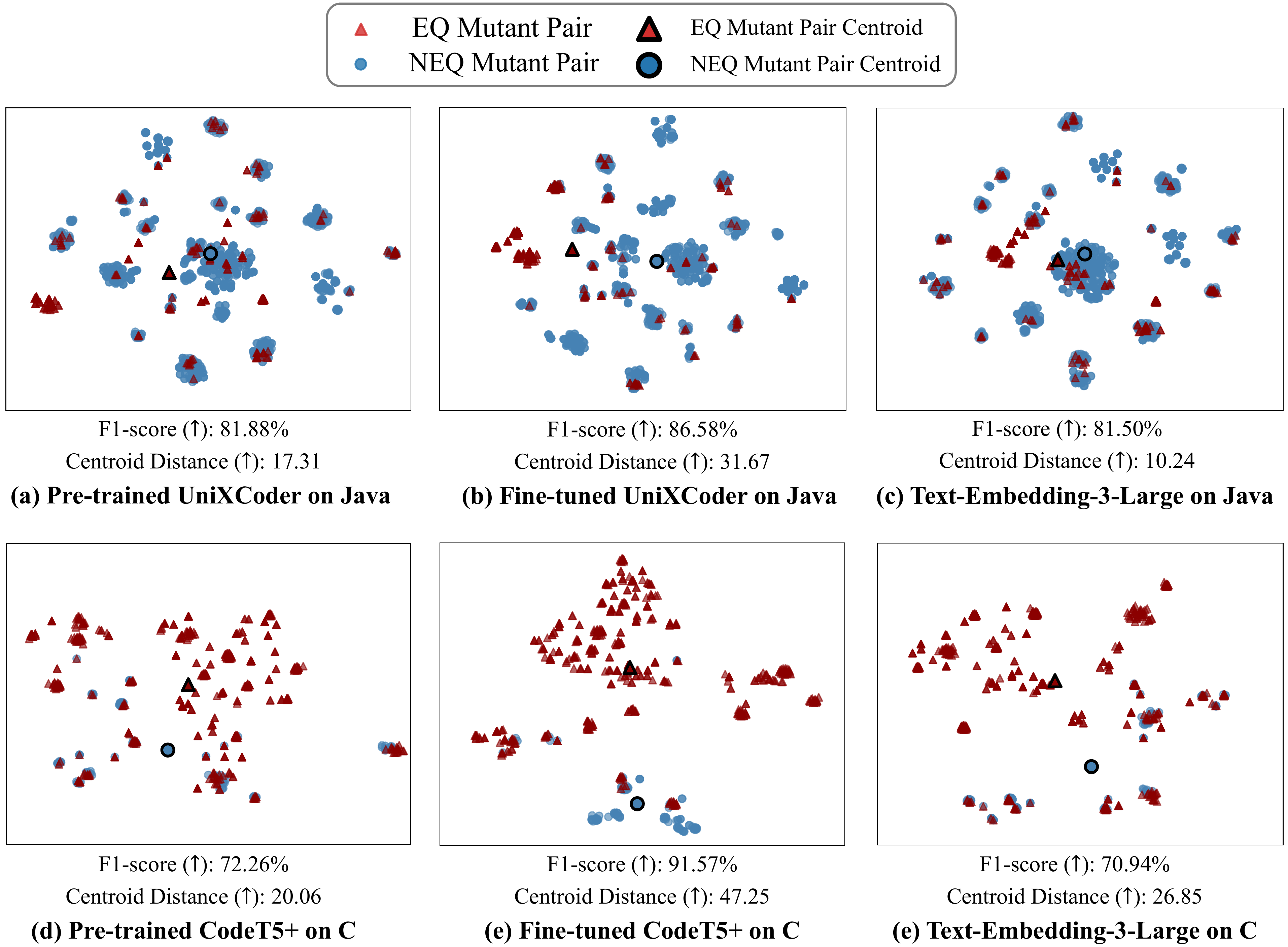}
    \caption{t-SNE plots showing the embedding of mutant pairs.
    EQ/NEQ represents equivalent/non-equivalent, respectively
    }
    \label{fig:embedding}
\end{figure}

\subsubsection{Does the embedding quality affect detection performance?}
Prior research~\cite{DBLP:conf/nips/LiTGLH22, DBLP:journals/jair/BurkartH21} demonstrates that embedding quality significantly impacts the ability to capture program semantic features for establishing effective decision boundaries.
To evaluate embedding quality of mutant pairs, we employed t-distributed stochastic neighbor embedding (t-SNE)~\cite{Maaten2008VisualizingDU}, which projects code embeddings into a 2-dimensional space for visual analysis of relationships between embeddings generated by various encoder LLMs.
Based on our findings from RQ1 and RQ2, we selected the top-performing LLMs from three categories: pre-trained code embedding, fine-tuned code embedding, and general text-embedding models. For Java, we selected pre-trained UniXCoder, fine-tuned UniXCoder, and Text-Embedding-3-Large; for C, we selected pre-trained CodeT5+, fine-tuned CodeT5+, and Text-Embedding-3-Large.
Following established methodology~\cite{ahmed2024studying}, we used \textit{centroid distance} to quantify embedding separation and quality. Higher centroid distance values indicate superior embedding quality and clearer delineation in the embedding space.

Figure~\ref{fig:embedding} presents t-SNE visualizations for all 1,650 Java mutant pairs and 544 C mutant pairs in our test set across the three selected LLMs. In the Java environment, fine-tuned UniXCoder achieved the best separation with a centroid distance of 31.67, outperforming both pre-trained UniXCoder (17.31) and Text-Embedding-3-Large (10.24).
Similarly, in the C environment, fine-tuned CodeT5+ demonstrated superior separation with a centroid distance of 47.25, compared to pre-trained CodeT5+ (20.06) and Text-Embedding-3-Large (26.85).
The strong correlation between centroid distance (embedding quality) and F1-score (detection performance) across all models confirms that embedding quality directly influences detection effectiveness.

\subsection{Qualitative Analysis of Incorrect LLM Detections}
\begin{figure*}[thp!]
    \centering
    \includegraphics[width=.95\linewidth]{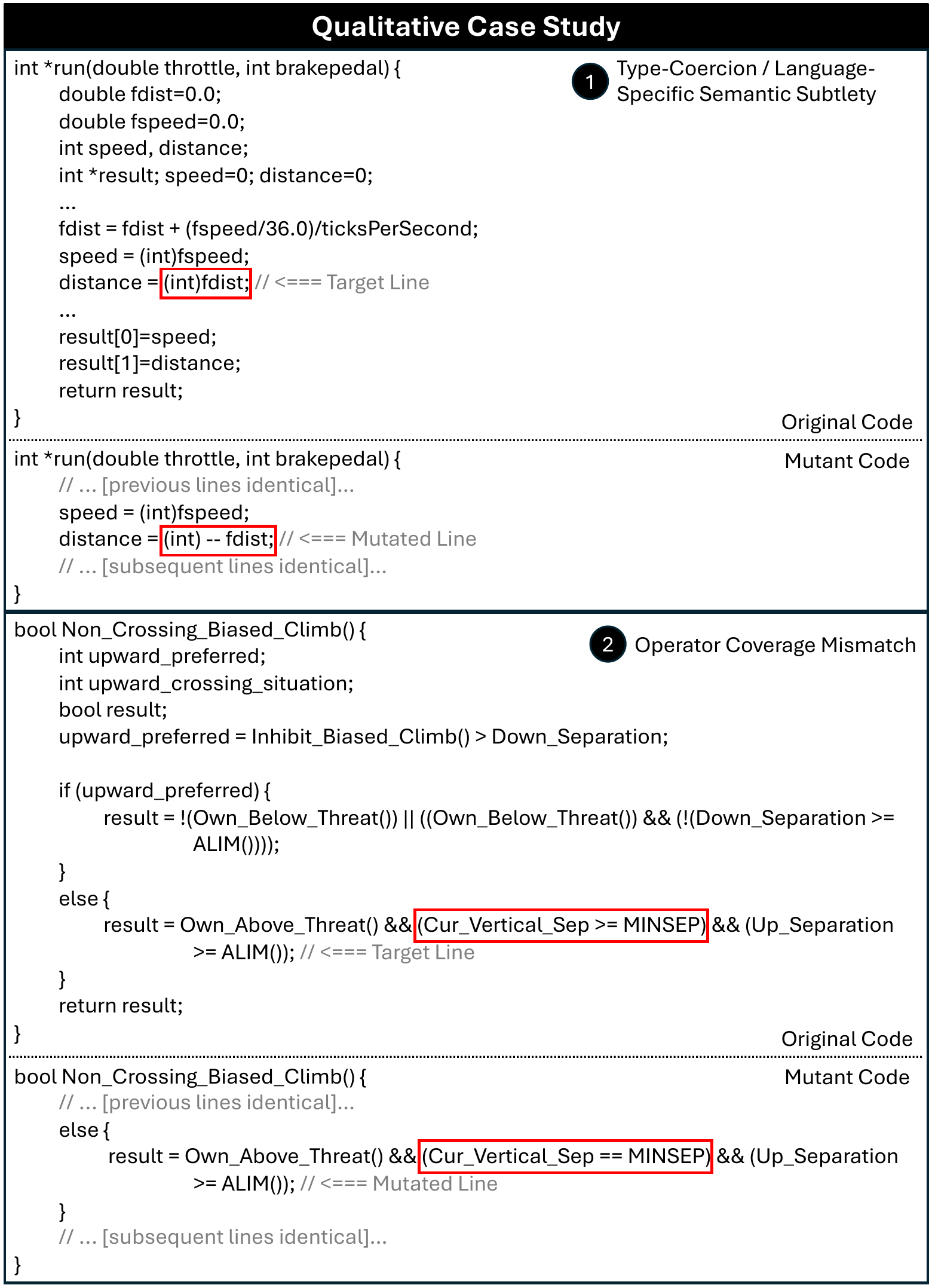}
    \caption{Case studies of incorrect recommendations}
    \label{fig:case_studies}
\end{figure*}

While our quantitative study demonstrates the effectiveness of LLMs in detecting equivalent mutants, aggregate evaluation metrics can obscure the nuanced trade-offs between different types of detection techniques.
Relying solely on these metrics overlooks the inherent limitations of LLMs and the unique advantages of conventional EMD techniques.
To provide a deeper understanding of why specific techniques excel in different scenarios, we conducted a rigorous manual analysis of the LLM's incorrect detections. This included analyzing both general failures and unique failures where the LLM failed, but conventional techniques succeeded.
Based on the results of RQ3, we analyzed a total of 129 incorrect detections (103 from the Java dataset and 26 from the C dataset). 
The first author manually annotated and classified each incorrect recommended mutant pair.
By analyzing the incorrect recommended mutant pairs across our C and Java datasets, we identified a clear taxonomy of LLM failure modes. 
The distribution reveals that LLM blind spots are heavily concentrated in structural and semantic subtleties:

\begin{enumerate}[label=\Roman*.]
    \item \textbf{Type-Coercion / Language-Specific Semantic Subtlety (37 instances)}: The failure stems from a C-specific operator or language rule, principally the side-effect semantics of pre/post increment ($++$) and pre/post decrement operators ($--$), where the LLM does not correctly model how the operator modifies a variable as a side effect within or across expressions.
    \item \textbf{Operator Coverage Mismatch (33 instances):} The mutant replaces a relational operator with one that has a different but partially overlapping coverage set, neither a complement nor a mirror, but a different cardinality of satisfying values. Requires the LLM to reason about which specific value ranges each operator covers.
    \item \textbf{Program Invariant / Pre-condition Blindness (32 instances):} The failure requires tracking a variable constraint or program state established by earlier statements within the same method. The LLM treats each statement in isolation rather than propagating intra-method invariants forward, causing it to miss either that the mutation is equivalent (given the constrained domain) or that it is non-equivalent (given a guaranteed invariant).
    \item \textbf{Subtle Semantic Equivalence via Algebraic Identity (19 instances):} The mutant pair is genuinely equivalent, but recognizing equivalence requires non-trivial algebraic or arithmetic reasoning, such as recognizing that two expressions produce identical values through different computational paths, or that a boundary case is neutralized by surrounding code.
    \item \textbf{Logical Complement Confusion (4 instances):} The mutant replaces a relational operator with its logical complement (e.g., from $>=$ to $<$), producing a condition that always yields the opposite boolean value. This is the most severe ROR failure. 
    \item \textbf{Symmetric Boundary Reversal (4 instances):} The mutant replaces a relational operator with its directional mirror (e.g., from $>=$ to $<=$). The two operators agree only at the exact equality point and disagree on both sides of it. Less severe than Logical Complement Confusion but still a clear semantic difference across most of the input domain.
\end{enumerate}

To illustrate how these inherent limitations contrast with the strengths of structural and execution-based techniques, we select two typical examples as the case studies.
As shown in Figure\mbox{~\ref{fig:case_studies}}(1), this mutant injects a pre-decrement operator `\texttt{$--$}' directly into a type-casting assignment. 
The LLM misclassified this as equivalent, treating `\texttt{fdist}' and `\texttt{$--$fdist}' as semantically similar within the context of the distance calculation. 
It failed to account for the side-effect: the pre-decrement permanently mutates the `\texttt{fdist}' variable's state in memory. 
Tree-based NN and ML-based techniques both failed here because the structural change is minimal.
However, the Compiler-based technique excelled, as the state mutation is immediately registered in the runtime execution environment, explicitly flagging the non-equivalence.
As shown in Figure\mbox{~\ref{fig:case_studies}}(2), in this collision avoidance logic, the relational operator `\texttt{>=}' is mutated to `\texttt{==}'. 
This mutation creates a severe Operator Coverage Mismatch. 
Mathematically, it drastically narrows the set of satisfying values from a continuous range (any vertical separation greater than or equal to the minimum threshold) to a single, exact boundary point.
A potential reason the LLM incorrectly detected this as an equivalent mutant is that the LLM lacks rigorous cardinality reasoning; the high textual similarity of the tokens around the operator caused it to hallucinate that the overall logic remained functionally equivalent.
Interestingly, the traditional ML-based and Compiler-based baseline techniques also failed to detect this mutation. 
The latter is likely due to the dynamic test suite lacking a specific boundary test where `\texttt{Cur\_Vertical\_Sep}' strictly exceeds `\texttt{MINSEP}'.

These case studies demonstrate that LLMs' semantic understanding cannot replace the structural rigor of conventional EMD techniques; relying solely on text-based attention is insufficient. 
Consequently, future research must shift away from treating LLMs as standalone solutions. 
Beyond a single solution, multi-stage hybrid pipelines offer immense potential. 
Compiler-based techniques can act as scalable initial filters to prune obvious equivalent mutants, routing ambiguous or structurally complex cases to LLM-based techniques for rigorous dynamic analysis.

\subsection{Sensitivity to the Treatment of Uncompilable Programs}
This discussion isolates the effect of uncompilable programs on compiler-based EMD techniques (i.e., TCE), which we regard as a property of the build environment rather than an intrinsic property of the techniques themselves. 
To quantify the impact of this issue, we evaluated two alternative strategies for handling uncompilable mutant pairs. 
The first is the \textbf{conservative classification strategy} adopted in our main experiments, under which every mutant pair that cannot be compiled is treated as an incorrect recommendation. 
This strategy reflects a strict deployment perspective, in which failure to produce a recommendation is considered a failure of the technique. 
The second is the \textbf{random classification strategy}, under which an uncompilable mutant pair is assigned the equivalent or non-equivalent label with equal probability (i.e., 50\% for each class). 
This strategy reflects the assumption that an uncompilable pair provides no information for determining semantic equivalence and therefore should not systematically penalize either class.

Table~\mbox{\ref{tab:rq6}} summarizes the results under the two strategies. 
The treatment of uncompilable pairs has a substantial and uneven impact on compiler-based techniques across programming languages. 
Under the conservative classification strategy, TCE$_{gcc}$, TCE$_{clang}$, TCE$_{Javac}$ and TCE$_{Soot}$ achieve the F1-scores of 10.08\%, 10.08\%, 39.31\% and 50.80\%, respectively.
Under the random classification strategy, the performance of the compiler-based techniques improves substantially. 
The effect is particularly pronounced for the C dataset. 
Specifically, the F1-score of both TCE$_{gcc}$ and TCE$_{clang}$ increases from 10.08\% to 47.17\%. 
Additionally, TCE$_{Javac}$ and TCE$_{Soot}$ also benefit from this alternative treatment, achieving the F1-scores of 49.49\% and 57.04\%.
Despite these changes in absolute performance, the relative ordering of the compiler-based techniques remains unchanged under the two classification strategies.
More importantly, the alternative treatment of uncompilable cases does not affect the overarching comparative findings or conclusions drawn from RQ1 to RQ5. 
These results suggest that, although the handling of uncompilable programs can substantially affect the absolute performance metrics of compiler-based EMD techniques, it does not undermine the main comparative conclusions of our study.

\setcounter{table}{5}
\begin{table}[thp!]
\caption{The performance analysis of uncompilable programs on compiler-based EMD techniques}
\label{tab:rq6}
\centering
    \begin{tabular}{lrrrrrr} 
    \toprule
    \toprule
    \multirow{2}{*}{\textbf{Technique}} & \multicolumn{3}{c}{\textbf{C}} & \multicolumn{3}{c}{\textbf{Java}} \\ \cmidrule(lr){2-4} \cmidrule(lr){5-7}
     & \textbf{Precision} & \textbf{Recall} & \textbf{F1-score} & \textbf{Precision} & \textbf{Recall} & \textbf{F1-score} \\ 
    \midrule
    \multicolumn{7}{l}{\framecolorbox[11.1cm][l]{gray!30}{gray!30}{\textbf{Compiler-based technique (conservative classification strategy)}}} \\ 
    \midrule
    TCE$_{gcc}$ & 5.93\% & 33.52\% & 10.08\% & -- & -- & --\\
    TCE$_{clang}$ & 5.93\% & 33.52\% & 10.08\% & -- & -- & -- \\
    TCE$_{Javac}$ & -- & -- & -- & 40.55\% & 38.15\% & 39.31\% \\
    TCE$_{Soot}$ & -- & -- & -- & 51.26\% & 51.81\% & 50.80\% \\
    \midrule
    \multicolumn{7}{l}{\framecolorbox[11.1cm][l]{gray!30}{gray!30}{\textbf{Compiler-based technique (random classification strategy)}}} \\ 
    \midrule
    TCE$_{gcc}$ & 57.01\% & 62.07\% & 47.17\% & -- & -- & --\\
    TCE$_{clang}$ & 57.01\% & 62.07\% & 47.17\% & -- & -- & -- \\
    TCE$_{Javac}$ & -- & -- & -- & 49.81\% & 49.75\% & 49.49\% \\
    TCE$_{Soot}$ & -- & -- & -- & 56.63\% & 59.54\% & 57.04\% \\
    \bottomrule
    \bottomrule
    \end{tabular}
\end{table}

\subsection{Future Work}
\subsubsection{Leveraging Reasoning Ability of LLMs.} 
In our study, we solely utilized standard prompting approaches (i.e., zero-shot and few-shot prompting).
Recent research has introduced Chain-of-Thought (CoT)~\cite{kojima2022large,tian2023test} prompting to activate reasoning capabilities in LLMs, alongside reasoning-enhanced LLMs~\cite{guo2025deepseek, chatgpt2025} developed through reinforcement learning.
These techniques help tackle complex problems (e.g., mathematical reasoning and code generation) by employing intermediate reasoning processes to derive solutions.
Multiple studies have confirmed CoT prompting's effectiveness in improving LLM performance across complex reasoning benchmarks~\cite{wei2022chain,jiang2023self}. Similarly, reasoning LLMs have demonstrated significant effectiveness on complex tasks.
Therefore, we could further investigate how CoT prompting or reasoning LLMs might improve equivalent mutant detection.

\subsubsection{Equivalent Mutant Avoidance.} 
In contrast to detection approaches, researchers have also explored strategies to avoid generating equivalent mutants entirely~\cite{madeyski2013overcoming}. 
These avoidance techniques employ sophisticated methods such as program dependence analysis and higher-order mutation operators to significantly reduce the occurrence of equivalent mutants~\cite{harman2001relationship,jia2009higher,oh2021effectively}. 
Interestingly, detection and avoidance techniques have largely developed as separate research streams. 
A promising direction for future work lies in integrating these complementary approaches to create more efficient and effective mutation testing frameworks.

\subsubsection{Duplicated Mutant Detection.} 
Mutant duplication poses a significant challenge in mutation testing. 
These duplicated mutants maintain semantic equivalence with other mutants while potentially differing from the original program semantically. 
Such duplications can misleadingly inflate a test suite's apparent effectiveness at killing mutants, compromising the validity of results.
~\citet{kintis2017detecting} has shown that TCE, initially developed for equivalent mutant detection, effectively identifies these duplicated mutants. 
Extending this work, our upcoming research will explore the capabilities of LLMs in detecting duplicated mutants, potentially providing more efficient and accurate solutions to this ongoing challenge in the field.

\subsubsection{Stubborn Mutant Detection.}
Our study deliberately focuses on the binary classification setting of distinguishing equivalent from non-equivalent mutants, allowing us to establish a foundational understanding of how LLMs reason about semantic equivalence.
In practice, however, an important intermediate category consists of stubborn mutants, semantically non-equivalent mutants but remain difficult to kill because identifying triggering inputs or execution paths is particularly non-trivial\mbox{~\cite{chekam2021killing}}. 
Distinguishing true equivalent mutants from stubborn mutants remains challenging due to the lack of large-scale datasets with reliable fine-grained labels. Such labeling requires not only establishing semantic equivalence, but also demonstrating that a mutant survives extensive testing efforts, substantially increasing annotation cost and complexity.
Addressing this challenge represents an important direction for future work. As high-quality datasets become available, the approaches investigated in this paper can be naturally extended beyond binary classification. For example, code-embedding models could be reformulated as multi-class classifiers that distinguish equivalent, stubborn, and readily killable mutants, while instruction-tuned LLMs could be leveraged to generate candidate execution paths or test-generation strategies for killing stubborn mutants. Such capabilities would move learning-based mutation analysis beyond classification and toward actionable testing assistance.

\section{Threats to Validity}
\label{sec:threats}
\textit{External Threats.} 
The main threat to validity lies in our equivalent mutant dataset. Our study focused only on Java and C programs, so our results may not generalize to other programming languages like Go. 
Additionally, our sample size is relatively small, which could introduce bias to our findings. 
In future work, we plan to expand our study framework to evaluate LLM performance across a wider range of programming languages and increase the number of samples in equivalent mutant detection.

\noindent
\textit{Internal Threats.} 
The primary threat lies in the implementation accuracy of each examined EMD technique. 
To address this threat, we implemented these techniques using the original open-source tools provided by the respective papers. 
Additionally, four authors conducted thorough code reviews to ensure implementation fidelity.
Another potential internal threat lies in our baseline evaluation of compiler-based approaches, which did not achieve the theoretical expectation of 100\% precision. 
This reduction does not stem from underlying compiler anomalies or bugs, but is instead an artifact of our strict evaluation protocol and metric computation. 
To evaluate practical deployability, we conservatively categorized compilation failures, caused by environmental and dependency inconsistencies, as incorrect recommendations. 
Furthermore, our use of macro-averaged precision on a class-imbalanced dataset mathematically amplified the impact of these uncompilable cases on the final metric. 
We mitigate this threat by conducting a reproduction of the standard, non-conservative evaluation protocol. When compilation failures are simply classified as non-equivalent mutants and evaluated using a binary-average metric, the approach achieves the theoretically expected 100\% precision, thereby validating the correctness of the binary-comparison mechanism itself.

\noindent
\textit{Construct Threats.} 
We acknowledge three related threats to the validity of our study.
First, potential bias may arise from our dataset construction methodology. While we employed stratified sampling with a 50\% split, a common approach in the field, alternative sampling strategies could yield different results.
Second, our EQ/NEQ ratio (17.80\% in the Java environment) is not perfectly representative of all real-world scenarios. However, it offers a more realistic perspective than the 50.00\% ratio commonly used in previous research~\cite{brito2020preliminary,peacock2021automatic,ma2023scope}. 
In future work, we plan to conduct a comprehensive investigation of LLM sensitivity across various ratios in more practical benchmarks.
Finally, computational and cost constraints prevented us from executing our EMD techniques multiple times to address potential variance and randomness. We encourage future researchers to repeat these experiments and report averaged results for more robust findings.

\section{Conclusion}
\label{sec:conclusion}
This work conducts a comprehensive empirical study investigating the effectiveness and efficiency of LLMs for equivalent mutant detection in Java and C programming languages.
We evaluate 12 state-of-the-art LLMs against ten established EMD techniques, analyzing various LLM strategies, assessing technique orthogonality, measuring computational overhead, and testing cross-language generalization capabilities.
Our findings demonstrate that LLM-based approaches achieve higher F1-scores than the evaluated baseline techniques across both Java and C, with fine-tuned code embedding strategies yielding the highest detection metrics among the tested approaches. 
Importantly, our analysis indicates that LLM-based techniques offer a practical trade-off between computational cost and effectiveness. 
Furthermore, fine-tuned LLMs demonstrate measurable cross-language generalization capabilities between Java and C equivalent mutant detection tasks.
This work opens several promising research directions, including extending equivalent mutant detection to additional programming languages, exploring the benefits of chain-of-thought prompting and advanced reasoning capabilities in LLMs, investigating potential synergies with equivalent mutant avoidance techniques, and applying LLMs to the related challenge of duplicated mutant detection.

\begin{acks}
This work was supported by National Natural Science Foundation of China (Grant Nos. 62322208, 12411530122), JSPS for the KAKENHI grants (JP25K22845), Japan Science and Technology Agency (JST) as part of Adopting Sustainable Partnerships for Innovative Research Ecosystem (ASPIRE), Grant Nos JPMJAP2415, and the Inamori Research Institute for Science for supporting Yasutaka Kamei via the InaRIS Fellowship. 
\end{acks}

\balance
\bibliographystyle{ACM-Reference-Format}
\bibliography{reference}

\end{document}